\documentclass[english]{article}
\usepackage[T1]{fontenc}
\usepackage[latin9]{inputenc}
\usepackage{geometry}
\usepackage{amsmath}
\usepackage{amsthm}
\usepackage{amssymb}
\usepackage{graphicx}
\usepackage{setspace}
\usepackage[authoryear]{natbib}
\makeatletter
  \theoremstyle{definition}
  \newtheorem{condition}{\protect\conditionname}
\theoremstyle{plain}
\newtheorem{thm}{\protect\theoremname}
  \theoremstyle{remark}
  \newtheorem{rem}{\protect\remarkname}
  \theoremstyle{plain}
  \newtheorem{prop}{\protect\propositionname}

\usepackage[bottom]{footmisc}

\newcommand{\Perp}{\perp \! \! \! \perp}
\usepackage{amsmath}
\usepackage{etoolbox}
\patchcmd{\quote}{\rightmargin}{\leftmargin 2em\rightmargin}{}{}

\makeatother

\usepackage{babel}
  \providecommand{\conditionname}{Condition}
  \providecommand{\propositionname}{Proposition}
  \providecommand{\remarkname}{Remark}
\providecommand{\theoremname}{Theorem}

\begin{document}

\title{On Heckits, LATE, and Numerical Equivalence\thanks{We thank Josh Angrist, David Card, James Heckman, Magne Mogstad, Parag
Pathak, Demian Pouzo, Raffaele Saggio, and Andres Santos for helpful
discussions.}\\
 \bigskip{}
 }

\author{Patrick Kline\\
 UC Berkeley and NBER\and Christopher R. Walters\\
 UC Berkeley and NBER\bigskip{}
 }

\date{October 2018\bigskip{}
 }
\maketitle
\begin{abstract}
Structural econometric methods are often criticized for being sensitive
to functional form assumptions. We study parametric estimators of
the local average treatment effect (LATE) derived from a widely used
class of latent threshold crossing models and show they yield LATE
estimates algebraically equivalent to the instrumental variables (IV)
estimator. Our leading example is Heckman's \citeyearpar{heckman_1979}
two-step (``Heckit'') control function estimator which, with two-sided
non-compliance, can be used to compute estimates of a variety of causal
parameters. Equivalence with IV is established for a semi-parametric
family of control function estimators and shown to hold at interior
solutions for a class of maximum likelihood estimators. Our results
suggest differences between structural and IV estimates often stem
from disagreements about the target parameter rather than from functional
form assumptions per se. In cases where equivalence fails, reporting
structural estimates of LATE alongside IV provides a simple means
of assessing the credibility of structural extrapolation exercises.

\bigskip{}

\end{abstract}
\newpage{}

\section{Introduction}

In a seminal paper, \citet{ai94} proposed an interpretation of the
instrumental variables (IV) estimand as a Local Average Treatment
Effect (LATE) \textendash{} an average effect for a subpopulation
of ``compliers'' compelled to change treatment status by an external
instrument. The plausibility and transparency of the conditions underlying
this interpretation are often cited as an argument for preferring
IV estimators to nonlinear estimators based on parametric models \citep{angristpischke2009,angrist_pischke_jep}.
On the other hand, LATE itself has been criticized as difficult to
interpret, lacking in policy relevance, and problematic for generalization
\citep{heckman_jhr_1997,deaton_2009,heckman_urzua_2010}. Adherents
of this view favor estimators motivated by joint models of treatment
choice and outcomes with structural parameters defined independently
of the instrument at hand.

This note develops some connections between IV and structural estimators
intended to clarify how the choice of estimator affects the conclusions
researchers obtain in practice. Our first result is that, in the familiar
binary instrument/binary treatment setting with imperfect compliance,
a wide array of structural ``control function'' estimators derived
from parametric threshold-crossing models yield LATE estimates numerically
identical to IV. Notably, this equivalence applies to appropriately
parameterized variants of Heckman's (\citeyear{heckman_1976,heckman_1979})
classic two-step (``Heckit'') estimator that are nominally predicated
on bivariate normality. Differences between structural and IV estimates
therefore stem in canonical cases entirely from disagreements about
the target parameter rather than from functional form assumptions.

After considering how this result extends to settings with instruments
taking multiple values, we probe its limits by examining some estimation
strategies where equivalence fails. First, we revisit a control function
estimator considered by \citet{lalonde_1986} and show that it produces
results identical to IV only under a symmetry condition on the estimated
probability of treatment. Next, we study an estimator motivated by
a selection model that violates the monotonicity condition of \citet{ai94}
and establish that it yields a LATE estimate different from IV, despite
fitting the same sample moments. Standard methods of introducing observed
covariates also break the equivalence of control function and IV estimators,
but we discuss a reweighting approach that ensures equivalence is
restored. We then consider full information maximum likelihood (FIML)
estimation of some generalizations of the textbook bivariate probit
model and show that this yields LATE estimates that coincide with
IV at interior solutions. However, FIML diverges from IV when the
likelihood is maximized on the boundary of the structural parameter
space, which serves as the basis of recent proposals for testing instrument
validity in just-identified settings \citep{huber_mellace,kitagawa_2015}.
Finally, we discuss why estimation of over-identified models generally
yields LATE estimates different from IV.

The equivalence results developed here provide a natural benchmark
for assessing the credibility of structural estimators, which typically
employ a number of over-identifying restrictions in practice. As \citet{angrist_pischke_jep}
note: ``A good structural model might tell us something about economic
mechanisms as well as causal effects. But if the information about
mechanisms is to be worth anything, the structural estimates should
line up with those derived under weaker assumptions.'' Comparing
the model-based LATEs implied by structural estimators with unrestricted
IV estimates provides a transparent assessment of how conclusions
regarding a common set of behavioral parameters are influenced by
the choice of estimator. A parsimonious structural estimator that
rationalizes a variety of IV estimates may reasonably be deemed to
have survived a ``trial by fire,'' lending some credibility to its
predictions.

\section{Two views of LATE}

We begin with a review of the LATE concept and its link to IV estimation.
Let $Y_{i}$ represent an outcome of interest for individual $i$,
with potential values $Y_{i}(1)$ and $Y_{i}(0)$ indexed against
a binary treatment $D_{i}$. Similarly, let $D_{i}(1)$ and $D_{i}(0)$
denote potential values of the treatment indexed against a binary
instrument $Z_{i}$. Realized treatments and outcomes are linked to
their potential values by the relations $D_{i}=Z_{i}D_{i}(1)+\left(1-Z_{i}\right)D_{i}(0)$
and $Y_{i}=D_{i}Y_{i}(1)+\left(1-D_{i}\right)Y_{i}(0)$. \citet{ai94}
consider instrumental variables estimation under the following assumptions:
\begin{enumerate}
\item[IA.1] Independence/Exclusion: $(Y_{i}(1),Y_{i}(0),D_{i}(1),D_{i}(0))\Perp Z_{i}$.
\item[IA.2] First Stage: $Pr\left[D_{i}=1|Z_{i}=1\right]>Pr\left[D_{i}=1|Z_{i}=0\right]$.
\item[IA.3] Monotonicity: $Pr\left[D_{i}(1)\geq D_{i}(0)\right]=1$.
\end{enumerate}
Assumption IA.1 requires the instrument to be as good as randomly
assigned and to influence outcomes only through its effect on $D_{i}$.
Assumption IA.2 requires the instrument to increase the probability
of treatment, and assumption IA.3 requires the instrument to weakly
increase treatment for all individuals.

\citet{ai94} define LATE as the average treatment effect for ``compliers''
induced into treatment by the instrument (for whom $D_{i}(1)>D_{i}(0))$.
Assumptions IA.1-IA.3 imply that the population \citet{wald_1940}
ratio identifies LATE:

\begin{center}
$\dfrac{E\left[Y_{i}|Z_{i}=1\right]-E\left[Y_{i}|Z_{i}=0\right]}{E\left[D_{i}|Z_{i}=1\right]-E\left[D_{i}|Z_{i}=0\right]}=E\left[Y_{i}(1)-Y_{i}(0)|D_{i}(1)>D_{i}(0)\right]\equiv LATE.$
\par\end{center}

Suppose we have access to an $iid$ vector of sample realizations
$\left\{ Y_{i},D_{i},Z_{i}\right\} _{i=1}^{n}$ obeying the following
condition:
\begin{condition}
\noindent \textbf{$\tfrac{1}{\sum_{i}Z_{i}}\sum_{i}Z_{i}D_{i}>\tfrac{1}{\sum_{i}(1-Z_{i})}\sum_{i}(1-Z_{i})D_{i}$.
\label{assu:1}}
\end{condition}
\noindent When IA.2 is satisfied the probability of Condition 1 being
violated approaches zero at an exponential rate in $n$. The analogy
principle suggests estimating LATE with:

\noindent 
\[
\widehat{LATE}^{IV}=\dfrac{\tfrac{1}{\sum_{i}Z_{i}}\sum_{i}Z_{i}Y_{i}-\tfrac{1}{\sum_{i}(1-Z_{i})}\sum_{i}(1-Z_{i})Y_{i}}{\tfrac{1}{\sum_{i}Z_{i}}\sum_{i}Z_{i}D_{i}-\tfrac{1}{\sum_{i}(1-Z_{i})}\sum_{i}(1-Z_{i})D_{i}}.
\]

\noindent This IV estimator is well-defined under Condition \ref{assu:1},
and is consistent for $LATE$ under assumptions IA.1-IA.3 and standard
regularity conditions.

\subsection*{Threshold-crossing representation}

\citet{vytlacil_2002} showed that the LATE model can be written as
a joint model of potential outcomes and self-selection in which treatment
is determined by a latent index crossing a threshold. Suppose treatment
status is generated by the equation

\begin{center}
$D_{i}=1\left\{ \psi(Z_{i})\geq V_{i}\right\} $,
\par\end{center}

\noindent where the latent variable $V_{i}$ is independently and
identically distributed according to some continuous distribution
with cumulative distribution function $F_{V}\left(.\right):\mathbb{R}\rightarrow[0,1]$,
and $\psi\left(.\right):\{0,1\}\rightarrow\mathbb{R}$ defines instrument-dependent
thresholds below which treatment ensues. Typically $F_{V}\left(.\right)$
is treated as a structural primitive describing a stable distribution
of latent costs and benefits influencing program participation that
exists independently of a particular instrument, as in the classic
selection models of \citet{roy_1951} and \citet{heckman_1974}. We
follow \citet{heckman_vytlacil_2005} and work with the equivalent
transformed model

\begin{equation}
D_{i}=1\left\{ P(Z_{i})\geq U_{i}\right\} ,\label{eq:choice}
\end{equation}

\noindent where $U_{i}\equiv F_{V}(V_{i})$ follows a uniform distribution
and $P(Z_{i})\equiv F_{V}(\psi(Z_{i}))$ is the propensity score.
The instrument $Z_{i}$ is presumed to increase the likelihood of
treatment ($P(1)>P(0)),$ and to be independent of $U_{i}$ and potential
outcomes:
\begin{equation}
(Y_{i}(1),Y_{i}(0),U_{i})\Perp Z_{i}.\label{eq:independence}
\end{equation}

The selection model defined by (\ref{eq:choice}) and (\ref{eq:independence})
is equivalent to the treatment effects model described by assumptions
IA.1-IA.3. Equation (\ref{eq:choice}) merely translates the behavioral
responses that are permitted in the LATE model into a partition of
the unit interval. In the terminology of \citet{air96}, assumption
IA.3 implies that the population consists of compliers with $D_{i}(1)>D_{i}(0)$,
``always takers'' with $D_{i}(1)=D_{i}(0)=1$, and ``never takers''
with $D_{i}(1)=D_{i}(0)=0$. The latent variable $U_{i}$ is defined
such that always takers have $U_{i}\in[0,P(0)]$, compliers have $U_{i}\in(P(0),P(1)]$,
and never takers have $U_{i}\in(P(1),1]$. Condition (\ref{eq:independence})
implies that potential outcomes and treatment choices are independent
of the instrument and imposes no further restrictions on the joint
distribution of these quantities. It follows that we can equivalently
define $LATE=E\left[Y_{i}(1)-Y_{i}(0)|P(0)<U_{i}\leq P(1)\right]$.

Though Vytlacil's \citeyearpar{vytlacil_2002} results establish equivalence
between a non-parametric latent index model and the LATE model, the
fully non-parametric model is typically not used for estimation. Rather,
to motivate alternatives to IV estimation, it is conventional to make
additional assumptions regarding the joint distribution of the latent
cost $U_{i}$ and the potential outcomes $\left(Y_{i}(1),Y_{i}(0)\right)$.
The goal of this note is to investigate the consequences of such assumptions
for empirical work.

\section{Control function estimation}

We begin by considering estimators predicated on the existence of
a parametric ``control function'' capturing the endogeneity in the
relationship between outcomes and treatment \citep{heckman_robb,blundell_matzkin,wooldridge_2015}.
The workhorse models in this literature obey the following semi-parametric
restriction:
\begin{equation}
E\left[Y_{i}(d)|U_{i}=u\right]=\alpha_{d}+\gamma_{d}\times\left(J(u)-\mu_{J}\right),\mbox{ \ensuremath{d\in\left\{ 0,1\right\} ,}}\ u\in(0,1),\label{eq:linear}
\end{equation}

\noindent where $J(\cdot):\left(0,1\right)\rightarrow\mathbb{R}$
is a strictly increasing continuous function and $\mu_{J}\equiv E\left[J(U_{i})\right]$.
\citet{lee_selection_1983} studied this dependence structure in the
context of classic ``one-sided'' selection problems where outcomes
are only observed when $D_{i}=1$. Setting $J(\cdot)$ equal to the
inverse normal CDF yields the canonical Heckman \citeyearpar{heckman_1976,heckman_1979}
sample selection (``Heckit'') model, while choosing $J(u)=u$ yields
the linear selection model studied by \citet{olsen_1980}, and choosing
the inverse logistic CDF for $J\left(.\right)$ yields the logit selection
model considered by \citet{mroz_1987}. 

Subsequent work applies versions of (\ref{eq:linear}) to policy evaluation
by modeling program participation as a ``two-sided'' sample selection
problem with coefficients indexed by the treatment state $d$. For
example, \citet{bjorklund_moffitt} build on the Heckit framework
by assuming $J(\cdot)$ is the inverse normal CDF and allowing $\alpha_{1}\neq\alpha_{0}$,
$\gamma_{1}\neq\gamma_{0}$. Likewise, the linear estimator of \citet{brinch_etal}
is a two-sided variant of Olsen's \citeyearpar{olsen_1980} approach
that imposes an identity $J(\cdot)$ function with coefficients indexed
by $d$. Interestingly, Dubin and McFadden's \citeyearpar{dubin_mcfadden}
classic multinomial selection model collapses in the binary treatment
effects case to a two-sided version of Mroz's \citeyearpar{mroz_1987}
logit model.

Assumption (\ref{eq:linear}) nullifies Vytlacil's \citeyearpar{vytlacil_2002}
equivalence result by imposing restrictions on the relationships between
mean potential outcomes of subgroups that respond differently to the
instrument $Z_{i}$. Let $\mu_{dg}$ denote the mean of $Y_{i}(d)$
for group $g\in\{at,nt,c\}$, representing always takers, never takers
and compliers. For any strictly increasing $J(\cdot)$, equation (\ref{eq:linear})
implies $sgn(\mu_{dat}-\mu_{dc})=sgn(\mu_{dc}-\mu_{dnt})$ for $d\in\{0,1\}$.
In contrast, the nonparametric model defined by assumptions IA.1-IA.3
is compatible with any arrangement of differences in mean potential
outcomes for the three subgroups. We next consider whether these additional
restrictions are consequential for estimation of LATE.

\subsection*{LATE}

When non-compliance is ``two-sided'' so that $0<P(0)<P(1)<1$, equation
(\ref{eq:linear}) implies that mean outcomes conditional on treatment
status are

\begin{center}
$E\left[Y_{i}|Z_{i},D_{i}=d\right]=\alpha_{d}+\gamma_{d}\lambda_{d}\left(P(Z_{i})\right)$,
\par\end{center}

\noindent where $\lambda_{1}(\cdot):\left(0,1\right)\rightarrow\mathbb{R}$
and $\lambda_{0}(\cdot):\left(0,1\right)\rightarrow\mathbb{R}$ are
control functions giving the means of $\left(J\left(U_{i}\right)-\mu_{J}\right)$
when $U_{i}$ is truncated from above and below at $p\in\left(0,1\right)$:

\begin{center}
$\lambda_{1}(p)=E\left[J(U_{i})-\mu_{J}|U_{i}\leq p\right],\ \lambda_{0}(p)=E\left[J(U_{i})-\mu_{J}|U_{i}>p\right]$.
\par\end{center}

\noindent \begin{flushleft}
While attention in parametric selection models often focuses on the
population average treatment effect $\alpha_{1}-\alpha_{0}$ \citep{garen_84,heckman_varieties,wooldridge_2015},
equation (\ref{eq:linear}) can also be used to compute treatment
effects for other subgroups. The average effect on compliers can be
written
\begin{equation}
LATE=\alpha_{1}-\alpha_{0}+\left(\gamma_{1}-\gamma_{0}\right)\Gamma\left(P(0),P(1)\right),\label{eq:complier_lambda}
\end{equation}
\par\end{flushleft}

\noindent where $\Gamma(p,p^{\prime})$ gives the mean of $J(U_{i})-\mu_{J}$
when $U_{i}$ lies between $p$ and $p^{\prime}>p$:

\begin{center}
$\Gamma(p,p^{\prime})=E\left[J(U_{i})-\mu_{J}|p<U_{i}\leq p^{\prime}\right]=\dfrac{p^{\prime}\lambda_{1}(p^{\prime})-p\lambda_{1}(p)}{p^{\prime}-p}$.
\par\end{center}

\noindent \begin{flushleft}
The last term in (\ref{eq:complier_lambda}) adjusts the average treatment
effect to account for non-random selection into compliance with the
instrument.
\par\end{flushleft}

\subsection*{Estimation}

To motivate control function estimation, suppose that the sample exhibits
two-sided non-compliance as follows:
\begin{condition}
\noindent $0<\sum_{i}1\{D_{i}=d\}Z_{i}<\sum_{i}1\left\{ D_{i}=d\right\} $
for $d\in\{0,1\}$.\label{assu:2}
\end{condition}
\noindent \begin{flushleft}
This condition requires at least one observation with every combination
of $Z_{i}$ and $D_{i}$. Condition \ref{assu:2} is satisfied with
probability approaching one at an exponential rate in $n$ whenever
$0<Pr[Z_{i}=1]<1$ and $0<P(z)<1$ for $z\in\{0,1\}$.
\par\end{flushleft}

Control function estimation typically proceeds in two steps, both
for computational reasons and because of the conceptual clarity of
plug-in estimation strategies \citep{heckman_1979,smith_blundell}.
Deferring a discussion of one-step estimation approaches to later
sections, we define the control function estimator as a procedure
which first fits the choice model in equation (\ref{eq:choice}) by
maximum likelihood, then builds estimates of $\lambda_{1}(\cdot)$
and $\lambda_{0}(\cdot)$ to include in second-step ordinary least
squares (OLS) regressions for each treatment category. The first step
estimates can be written
\begin{equation}
\left(\hat{P}(0),\hat{P}(1)\right)={\displaystyle \arg\max_{P(0),P(1)}\sum_{i}D_{i}\log P(Z_{i})+\sum_{i}(1-D_{i})\log\left(1-P(Z_{i})\right)}.\label{eq:mle}
\end{equation}

\noindent The second step OLS estimates are
\begin{equation}
\left(\hat{\alpha}_{d},\hat{\gamma}_{d}\right)={\displaystyle \arg\min_{\alpha_{d},\gamma_{d}}}\sum_{i}1\left\{ D_{i}=d\right\} \left[Y_{i}-\alpha_{d}-\gamma_{d}\lambda_{d}(\hat{P}(Z_{i}))\right]^{2},\ d\in\{0,1\}.\label{eq:second_step}
\end{equation}
The analogy principle then suggests the following plug-in estimator
of LATE:

\begin{center}
$\widehat{LATE}^{CF}=\left(\hat{\alpha}_{1}-\hat{\alpha}_{0}\right)+\left(\hat{\gamma}_{1}-\hat{\gamma}_{0}\right)\Gamma(\hat{P}(0),\hat{P}(1))$.
\par\end{center}

Note that when non-compliance is ``one-sided'' so that $\sum_{i}D_{i}(1-Z_{i})=0$
or $\sum_{i}(1-D_{i})Z_{i}=0$, the maximum likelihood estimates in
(\ref{eq:mle}) are not well-defined. Condition  \ref{assu:2} ensures
that $\hat{P}(0)$ and $\hat{P}(1)$ exist, and that $\hat{\alpha}_{d}$
and $\hat{\gamma}_{d}$ can be computed for each value of $d$. Condition
\ref{assu:1} additionally ensures that $\hat{P}(0)<\hat{P}(1)$,
guaranteeing that $\widehat{LATE}^{CF}$ exists.

\section{\label{sec:Equivalence}Equivalence results}

Compared to $\widehat{LATE}^{IV}$, $\widehat{LATE}^{CF}$ would seem
to be highly dependent upon the functional form assumed for $J(\cdot)$
and the linearity of equation (\ref{eq:linear}). Our first result
shows that this is not the case.
\begin{thm}
\noindent If Conditions \ref{assu:1} and \ref{assu:2} hold then
$\widehat{LATE}^{CF}=\widehat{LATE}^{IV}$.\label{thm:LATE_equivalence}
\end{thm}
\noindent \textbf{Proof:} The maximum likelihood procedure in (\ref{eq:mle})
yields the empirical treatment rates $\hat{P}(z)=\tfrac{\sum_{i}1\left\{ Z_{i}=z\right\} D_{i}}{\sum_{i}1\{Z_{i}=z\}}$
for $z\in\{0,1\}$. The second-step OLS regressions can be rewritten

\noindent \begin{center}
$\left(\hat{\alpha}_{d},\hat{\gamma}_{d}\right)={\displaystyle \arg\min_{\alpha_{d},\gamma_{d}}}\sum_{i}1\left\{ D_{i}=d\right\} \left(Y_{i}-\left[\alpha_{d}+\gamma_{d}\lambda_{d}(\hat{P}(0))\right]-\gamma_{d}\left[\lambda_{d}(\hat{P}(1))-\lambda_{d}(\hat{P}(0))\right]Z_{i}\right)^{2}$.
\par\end{center}

\noindent This is a least squares fit of $Y_{i}$ on an intercept
and the indicator $Z_{i}$ in the subsample with $D_{i}=d$. Such
regressions can be estimated as long as there is two-sided non-compliance
with the instrument $Z_{i}$, which follows from Condition \ref{assu:2}.
Defining $\bar{Y}_{d}^{z}\equiv\tfrac{\sum_{i}1\left\{ D_{i}=d\right\} 1\left\{ Z_{i}=z\right\} Y_{i}}{\sum_{i}1\left\{ D_{i}=d\right\} 1\left\{ Z_{i}=z\right\} }$,
we have

\begin{center}
$\bar{Y}_{d}^{0}=\hat{\alpha}_{d}+\hat{\gamma}_{d}\lambda_{d}(\hat{P}(0))$,
$\bar{Y}_{d}^{1}-\bar{Y}_{d}^{0}=\hat{\gamma}_{d}\left[\lambda_{d}(\hat{P}(1))-\lambda_{d}(\hat{P}(0))\right]$.
\par\end{center}

\noindent Under Condition \ref{assu:1}, we have $\lambda_{d}(\hat{P}(1))\neq\lambda_{d}(\hat{P}(0))$,
and this pair of equations can be solved for $\hat{\gamma}_{d}$ and
$\hat{\alpha}_{d}$ as

\begin{center}
$\hat{\gamma}_{d}=\tfrac{\bar{Y}_{d}^{1}-\bar{Y}_{d}^{0}}{\lambda_{d}(\hat{P}(1))-\lambda_{d}(\hat{P}(0))},\ \hat{\alpha}_{d}=\tfrac{\lambda_{d}(\hat{P}(1))\bar{Y}_{d}^{0}-\lambda_{d}(\hat{P}(0))\bar{Y}_{d}^{1}}{\lambda_{d}(\hat{P}(1))-\lambda_{d}(\hat{P}(0))}$.
\par\end{center}

We can therefore rewrite the control function estimate of LATE as

\begin{center}
$\widehat{LATE}^{CF}=\left(\left[\tfrac{\lambda_{1}(\hat{P}(1))\bar{Y}_{1}^{0}-\lambda_{1}(\hat{P}(0))\bar{Y}_{1}^{1}}{\lambda_{1}(\hat{P}(1))-\lambda_{1}(\hat{P}(0))}\right]-\left[\tfrac{\lambda_{0}(\hat{P}(1))\bar{Y}_{0}^{0}-\lambda_{0}(\hat{P}(0))\bar{Y}_{0}^{1}}{\lambda_{0}(\hat{P}(1))-\lambda_{0}(\hat{P}(0))}\right]\right)$
\par\end{center}

\begin{center}
$+\left(\left[\tfrac{\bar{Y}_{1}^{1}-\bar{Y}_{1}^{0}}{\lambda_{1}(\hat{P}(1))-\lambda_{1}(\hat{P}(0))}\right]-\left[\tfrac{\bar{Y}_{0}^{1}-\bar{Y}_{0}^{0}}{\lambda_{0}(\hat{P}(1))-\lambda_{0}(\hat{P}(0))}\right]\right)\times\left(\frac{\hat{P}(1)\lambda_{1}(\hat{P}(1))-\hat{P}(0)\lambda_{1}(\hat{P}(0))}{\hat{P}(1)-\hat{P}(0)}\right)$.
\par\end{center}

\noindent Using the fact that $\lambda_{0}(p)=-\lambda_{1}(p)p/(1-p)$,
this simplifies to

\begin{center}
$\widehat{LATE}^{CF}=\dfrac{\left[\hat{P}(1)\bar{Y}_{1}^{1}+(1-\hat{P}(1))\bar{Y}_{0}^{1}\right]-\left[\hat{P}(0)\bar{Y}_{1}^{0}+(1-\hat{P}(0))\bar{Y}_{0}^{0}\right]}{\hat{P}(1)-\hat{P}(0)}$,
\par\end{center}

\noindent which is $\widehat{LATE}^{IV}$. $\blacksquare$

\begin{rem}
An immediate consequence of Theorem \ref{thm:LATE_equivalence} is
that $\widehat{LATE}^{CF}$ is also equivalent to the coefficient
on $D_{i}$ associated with a least squares fit of $Y_{i}$ to $D_{i}$
and a first stage residual $D_{i}-\hat{P}(Z_{i})$. \citet{blundell_matzkin}
attribute the first proof of the equivalence between this estimator
and IV to \citet{telser_1964}. 
\end{rem}

\begin{rem}
Theorem \ref{thm:LATE_equivalence} extends the analysis of \citet{brinch_etal}
who observe that linear control function estimators produce LATE estimates
numerically equivalent to IV. The above result implies that a wide
class of non-linear control function estimators share this property.
With a binary treatment and instrument, an instrumental variables
estimate can always be viewed as the numerical output of a variety
of parametric control function estimators. 
\end{rem}

\subsection*{Potential outcome means}

Corresponding equivalence results hold for estimators of other parameters
identified in the LATE framework. \citet{imbens_rubin_97} and \citet{abadie_2002}
discuss identification and estimation of the treated outcome distribution
for always takers, the untreated distribution for never takers, and
both marginal distributions for compliers. Nonparametric estimators
of the four identified marginal mean potential outcomes are given
by

\begin{center}
$\hat{\mu}_{1at}^{IV}=\bar{Y}_{1}^{0}$, $\hat{\mu}_{0nt}^{IV}=\bar{Y}_{0}^{1}$,
\par\end{center}

\begin{center}
$\hat{\mu}_{1c}^{IV}=\tfrac{\hat{P}(1)\bar{Y}_{1}^{1}-\hat{P}(0)\bar{Y}_{1}^{0}}{\hat{P}(1)-\hat{P}(0)}$,
$\hat{\mu}_{0c}^{IV}=\tfrac{(1-\hat{P}(0))\bar{Y}_{0}^{0}-(1-\hat{P}(1))\bar{Y}_{0}^{1}}{\hat{P}(1)-\hat{P}(0)}$.
\par\end{center}

The corresponding control function estimators are:

\begin{center}
$\hat{\mu}_{1at}^{CF}=\hat{\alpha}_{1}+\hat{\gamma}_{1}\lambda_{1}(\hat{P}(0))$,
$\hat{\mu}_{0nt}^{CF}=\hat{\alpha}_{0}+\hat{\gamma}_{0}\lambda_{0}(\hat{P}(1))$,
\par\end{center}

\begin{center}
$\hat{\mu}_{dc}^{CF}=\hat{\alpha}_{d}+\hat{\gamma}_{d}\Gamma(\hat{P}(0),\hat{P}(1)),\ d\in\{0,1\}$.
\par\end{center}

\noindent The following proposition shows that these two estimation
strategies produce algebraically identical results.
\begin{prop}
\noindent If Conditions \ref{assu:1} and \ref{assu:2} hold then\label{prop:po_means}
\end{prop}
\begin{center}
$\hat{\mu}_{dg}^{CF}=\hat{\mu}_{dg}^{IV}$ for $(d,g)\in\{(1,at),(0,nt),(1,c),(0,c)\}$.
\par\end{center}

\noindent \textbf{Proof:} Using the formulas from the proof of Theorem
\ref{thm:LATE_equivalence}, the control function estimate of $\mu_{1at}$
is

\begin{center}
$\hat{\mu}_{1at}^{CF}=\left(\tfrac{\lambda_{1}(\hat{P}(1))\bar{Y}_{1}^{0}-\lambda_{1}(\hat{P}(0))\bar{Y}_{1}^{1}}{\lambda_{1}(\hat{P}(1))-\lambda_{1}(\hat{P}(0))}\right)+\left(\tfrac{\bar{Y}_{1}^{1}-\bar{Y}_{1}^{0}}{\lambda_{1}(\hat{P}(1))-\lambda_{1}(\hat{P}(0))}\right)\lambda_{1}(\hat{P}(0))=\bar{Y}_{1}^{0}$,
\par\end{center}

\noindent which is $\hat{\mu}_{1at}^{IV}$. Likewise, 

\begin{center}
$\hat{\mu}_{0nt}^{CF}=\left(\tfrac{\lambda_{0}(\hat{P}(1))\bar{Y}_{0}^{0}-\lambda_{0}(\hat{P}(0))\bar{Y}_{1}^{1}}{\lambda_{0}(\hat{P}(1))-\lambda_{0}(\hat{P}(0))}\right)+\left(\tfrac{\bar{Y}_{0}^{1}-\bar{Y}_{0}^{0}}{\lambda_{0}(\hat{P}(1))-\lambda_{0}(\hat{P}(0))}\right)\lambda_{0}(\hat{P}(1))=\bar{Y}_{0}^{1}$
,
\par\end{center}

\noindent which is $\hat{\mu}_{0nt}^{IV}$. The treated complier mean
estimate is

\begin{center}
$\hat{\mu}_{1c}^{CF}=\left(\tfrac{\lambda_{1}(\hat{P}(1))\bar{Y}_{1}^{0}-\lambda_{1}(\hat{P}(0))\bar{Y}_{1}^{1}}{\lambda_{1}(\hat{P}(1))-\lambda_{1}(\hat{P}(0))}\right)+\left(\tfrac{\bar{Y}_{1}^{1}-\bar{Y}_{1}^{0}}{\lambda_{1}(\hat{P}(1))-\lambda_{1}(\hat{P}(0))}\right)\times\left(\frac{\hat{P}(1)\lambda_{1}(\hat{P}(1))-\hat{P}(0)\lambda_{1}(\hat{P}(0))}{\hat{P}(1)-\hat{P}(0)}\right)$
\par\end{center}

\begin{center}
$=\tfrac{\left(\lambda_{1}(\hat{P}(1))-\lambda_{1}(\hat{P}(0))\right)\hat{P}(1)\bar{Y}_{1}^{1}-\left(\lambda_{1}(\hat{P}(1))-\lambda_{1}(\hat{P}(0))\right)\hat{P}(0)\bar{Y}_{1}^{0}}{\left(\lambda_{1}(\hat{P}(1))-\lambda_{1}(\hat{P}(0))\right)\left(\hat{P}(1)-\hat{P}(0)\right)}=\tfrac{\hat{P}(1)\bar{Y}_{1}^{1}-\hat{P}(0)\bar{Y}_{1}^{0}}{\hat{P}(1)-\hat{P}(0)}$,
\par\end{center}

\noindent which is $\hat{\mu}_{1c}^{IV}$. Noting that $\widehat{LATE}^{IV}=\hat{\mu}_{1c}^{IV}-\hat{\mu}_{0c}^{IV}$
and $\widehat{LATE}^{CF}=\hat{\mu}_{1c}^{CF}-\hat{\mu}_{0c}^{CF}$,
it then follows by Theorem \ref{thm:LATE_equivalence} that $\hat{\mu}_{0c}^{CF}=\hat{\mu}_{0c}^{IV}$.
$\blacksquare$

\section{Equivalence and extrapolation}

Proposition \ref{prop:po_means} establishes that all control function
estimators based on equation (\ref{eq:linear}) produce identical
estimates of the potential outcome means that are nonparametrically
identified in the LATE framework. Different functional form assumptions
generate different estimates of quantities that are under-identified,
however. For example, the choice of $J(\cdot)$ in equation (\ref{eq:linear})
determines the shapes of the curves that the model uses to extrapolate
from estimates of the four identified potential outcome means $(\mu_{1at},\mu_{0nt},\mu_{1c},\mu_{0c})$
to the two under-identified potential outcome means $(\mu_{0at},\mu_{1nt})$.

\begin{center}
\includegraphics[scale=1.02]{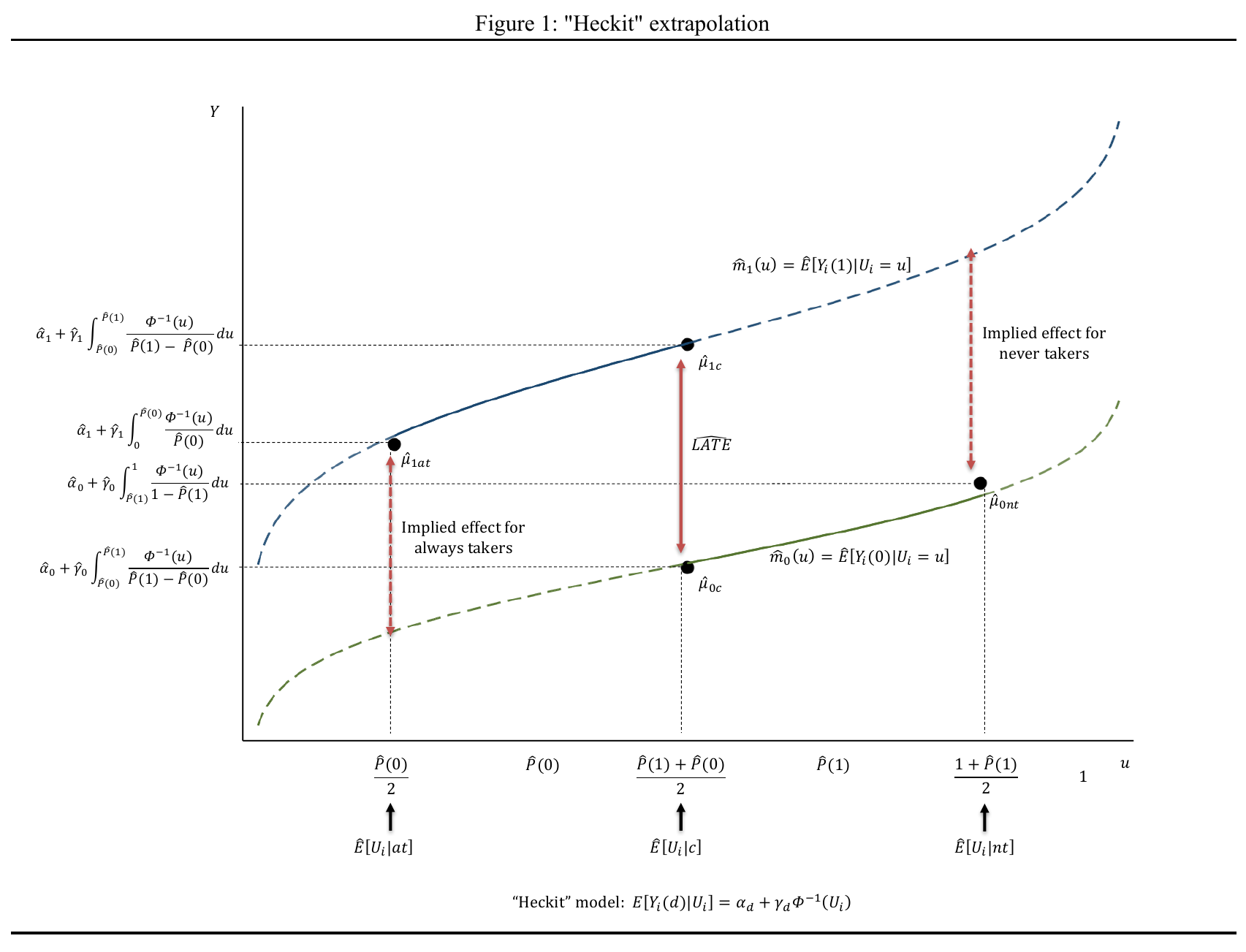}
\par\end{center}

Figures 1 and 2 illustrate this extrapolation in a hypothetical example.
The horizontal axis plots values $u$ of the unobserved treatment
cost $U_{i}$, while the vertical axis plots mean potential outcomes
$m_{d}(u)=E\left[Y_{i}(d)|U_{i}=u\right]$ as functions of this cost.
Estimates of these functions are denoted $\hat{m}_{d}(u)=\hat{\alpha}_{d}+\hat{\gamma}_{d}\times(J(u)-\mu_{J})$
and their difference $\hat{m}_{1}\left(u\right)-\hat{m}_{0}\left(u\right)$
provides an estimate of the marginal treatment effect \citep{bjorklund_moffitt,heckman_vytlacil_2005,heckman_essential_heterogeneity}
for an individual with latent cost $u$. 

Assumptions IA.1-IA3 ensure two averages of $m_{d}\left(U_{i}\right)$
are identified for each potential outcome: the treated means for always
takers and compliers, and the untreated means for never takers and
compliers. The control function estimator chooses $\hat{\alpha}_{d}$
and $\hat{\gamma}_{d}$ so that averages of $\hat{m}_{d}(U_{i})$
over the relevant ranges match the corresponding nonparametric estimates
for each compliance group. The coefficient $\hat{\gamma}_{1}$ parameterizes
the difference in mean treated outcomes between compliers and always
takers, while $\hat{\gamma}_{0}$ measures the difference in mean
untreated outcomes between compliers and never takers. Several tests
of endogeneous treatment assignment (see, e.g., \citealp{angrist_2004_tehet,battistin_rettore,berentha_imbens};
and \citealp{kowalski_2016}) amount to testing whether $\left(\hat{\gamma}_{0},\hat{\gamma}_{1}\right)$
are significantly different from zero. 

\begin{center}
\includegraphics[scale=0.98]{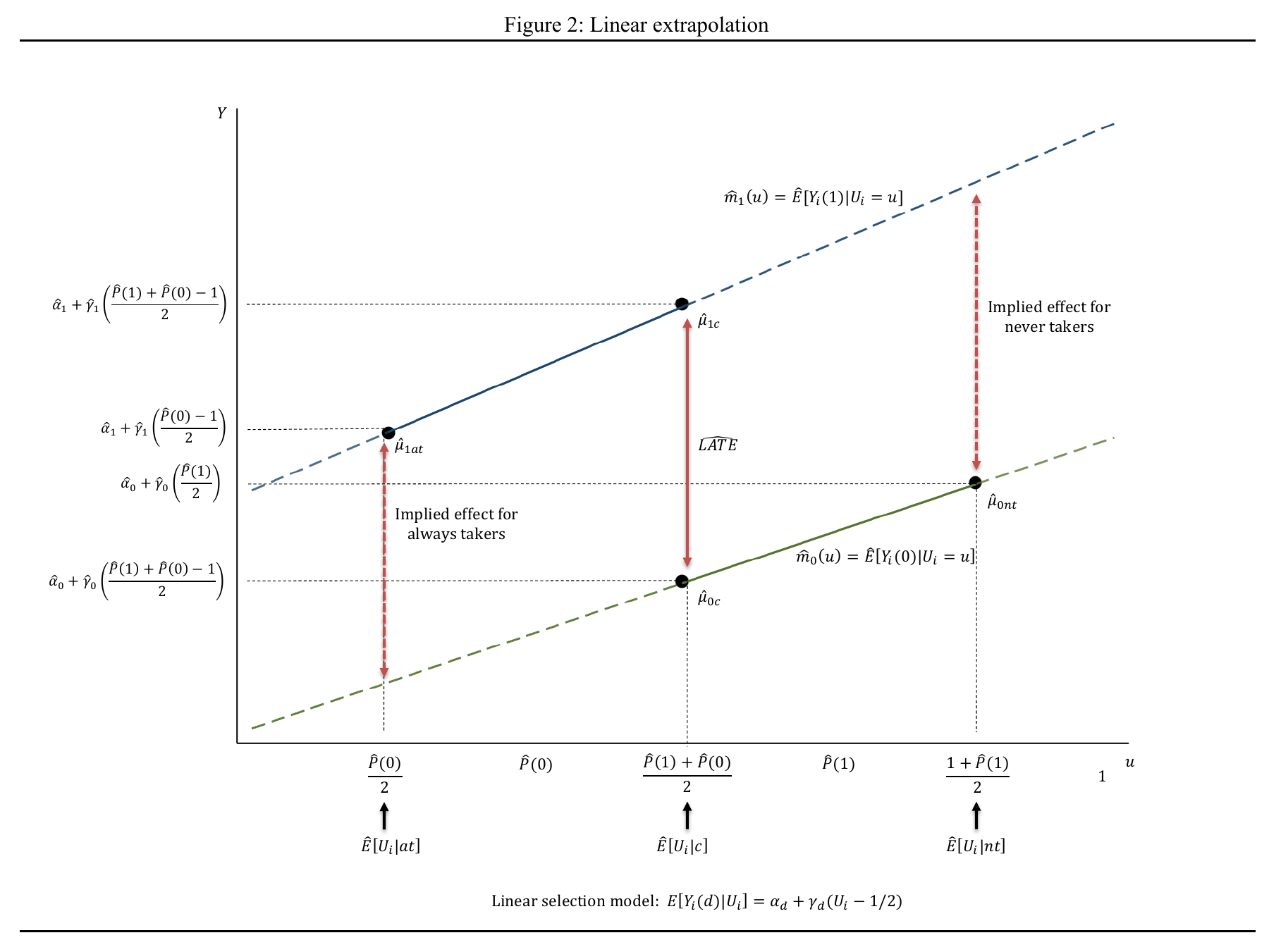}
\par\end{center}

Figure 1 depicts the results of parametric extrapolation based on
the Heckit model, while Figure 2 shows results for the linear control
function model discussed by \citet{brinch_etal}. Both models match
the same four estimated mean potential outcomes, thereby generating
identical estimates of LATE. Note that by Jensen's inequality, the
nonlinear $\hat{m}_{d}(u)$ curves in Figure 1 do not pass directly
through the group mean potential outcomes. The two models yield different
imputations for the missing potential outcomes of always takers and
never takers, and therefore also different estimates of the ATE, which
averages over all three subpopulations. This sensitivity to functional
form is intuitive: treatment effects for always and never takers are
fundamentally under-identified, an insight that has led to consideration
of bounds on these quantities \citep{manski_1990,balke_pearle,mst}.

\section{Multi-valued instruments}

Consider an instrument $Z_{i}$ taking values in $\{0,1,..,K\},$
and suppose that $0<\hat{P}\left(z-1\right)<\hat{P}\left(z\right)<1$
for $z\in\left\{ 1,2,...,K\right\} $. Let $D_{i}(z)$ denote $i$'s
treatment choice when $Z_{i}=z$. If assumptions IA.1-IA.3 hold for
every pair of instrument values, Wald ratios of the form $\frac{E\left[Y_{i}|Z_{i}=z\right]-E\left[Y_{i}|Z_{i}=z-1\right]}{E\left[D_{i}|Z_{i}=z\right]-E\left[D_{i}|Z_{i}=z-1\right]}$
identify the average treatment effect among compliers indexed by a
unit increment in the instrument, which we denote $LATE_{z}\equiv E\left[Y_{i}(1)-Y_{i}(0)|D_{i}(z)>D_{i}\left(z-1\right)\right]$.
Analog estimators of $LATE_{z}$ are given by the following pairwise
IV estimator: 
\[
\widehat{LATE}_{z}^{IV}=\dfrac{\tfrac{1}{\sum_{i}1\{Z_{i}=z\}}\sum_{i}1\{Z_{i}=z\}Y_{i}-\tfrac{1}{\sum_{i}1\{Z_{i}=z-1\}}\sum_{i}1\{Z_{i}=z-1\}Y_{i}}{\tfrac{1}{\sum_{i}1\{Z_{i}=z\}}\sum_{i}1\{Z_{i}=z\}D_{i}-\tfrac{1}{\sum_{i}1\{Z_{i}=z-1\}}\sum_{i}1\{Z_{i}=z-1\}D_{i}}.
\]
From Theorem \ref{thm:LATE_equivalence}, $\widehat{LATE}_{z}^{IV}$
is numerically equivalent to the corresponding pairwise control function
estimator of $LATE_{z}$ constructed from observations with $Z_{i}\in\{z-1,z\}$.
However, to improve precision, it is common to impose additional restrictions
on the $LATE_{z}$. 

Consider the following restriction on potential outcomes:
\begin{equation}
E\left[Y_{i}(d)|U_{i}\right]=\alpha_{d}+{\displaystyle \sum_{\ell=1}^{L}\gamma_{d\ell}\times\left(J(U_{i})-\mu_{J}\right)^{\ell}},\ d\in\{0,1\}.\label{eq:multi_z-1}
\end{equation}

\noindent Polynomial models of this sort have been considered by,
among others, \citet{brinch_etal} and \citet{dustmann_reverse_roy}.
Letting $\lambda_{1\ell}(p)=E\left[(J(U_{i})-\mu_{J})^{\ell}|U_{i}\leq p\right]$
and $\lambda_{0\ell}(p)=E\left[(J(U_{i})-\mu_{J})^{\ell}|U_{i}>p\right]$,
a two-step control function estimator of the parameters of equation
(\ref{eq:multi_z-1}) is

\noindent \begin{center}
$\left(\hat{\alpha}_{d},\hat{\gamma}_{d1},...,\hat{\gamma}_{dL}\right)={\displaystyle \arg\min_{\alpha_{d},\gamma_{d1},...,\gamma_{dL}}}\sum_{i}1\left\{ D_{i}=d\right\} \left[Y_{i}-\alpha_{d}-\sum_{\ell=1}^{L}\gamma_{d\ell}\lambda_{d\ell}(\hat{P}(Z_{i}))\right]^{2}$.
\par\end{center}

\noindent The resulting control function estimator of $LATE_{z}$
is then
\begin{equation}
\widehat{LATE}_{z}^{CF}=(\hat{\alpha}_{1}-\hat{\alpha}_{0})+\sum_{\ell=1}^{L}(\hat{\gamma}_{1\ell}-\hat{\gamma}_{0\ell})\Gamma_{\ell}(\hat{P}(z-1),\hat{P}(z)),\label{eq:multi_z_cf}
\end{equation}
where $\Gamma_{\ell}(p,p^{\prime})=[p^{\prime}\lambda_{1\ell}(p^{\prime})-p\lambda_{1\ell}(p)]/[p^{\prime}-p].$
The following proposition establishes that this estimator is identical
to $\widehat{LATE}_{z}^{IV}$ when $L=K$.
\begin{prop}
\noindent If Conditions 1 and 2 hold for every pair of instrument
values and the polynomial order $L$ equals $K$ then $\widehat{LATE}_{z}^{CF}=\widehat{LATE}_{z}^{IV}\ \forall z\in\{1,2,...,K\}$.\label{prop:multi_z_equivalence}
\end{prop}
\noindent \textbf{Proof:} See the Appendix. $\blacksquare$
\begin{rem}
\noindent Instrumenting $D_{i}$ with a scalar function $g(Z_{i})$
generates an IV estimate equal to a convex weighted average of the
$\widehat{LATE}_{z}^{IV}$ \citep{ai94}. From Proposition \ref{prop:multi_z_equivalence},
applying these weights to the $\widehat{LATE}_{z}^{CF}$ when $L=K$
will yield an identical result. By contrast, the set of $\widehat{LATE}_{z}^{CF}$
that result from imposing $L<K$ need not correspond to weighted averages
of the $\widehat{LATE}_{z}^{IV}$, but are likely to exhibit reduced
sampling variability.
\end{rem}

\begin{rem}
When $L<K-1$, the restriction in (\ref{eq:multi_z-1}) can be used
to motivate estimators of particular LATEs that are convex combinations
of IV estimators. In the case where $K=3$ and $L=1$, one can show
that:
\[
LATE_{2}=\left(\tfrac{\Gamma(P(1),P(2))-\Gamma(P(0),P(1))}{\Gamma(P(2),P(3))-\Gamma(P(0),P(1))}\right)LATE_{3}+\left(\tfrac{\Gamma(P(2),P(3))-\Gamma(P(1),P(2))}{\Gamma(P(2),P(3))-\Gamma(P(0),P(1))}\right)LATE_{1}.
\]
\end{rem}

\noindent This representation suggests combination estimators of the
form
\[
\widehat{LATE}{}_{2}^{\xi}=\xi\widehat{LATE}_{2}^{IV}+\left(1-\xi\right)\left[\left(\tfrac{\Gamma(\hat{P}(1),\hat{P}(2))-\Gamma(\hat{P}(0),\hat{P}(1))}{\Gamma(\hat{P}(2),\hat{P}(3))-\Gamma(\hat{P}(0),\hat{P}(1))}\right)\widehat{LATE}_{3}^{IV}+\left(\tfrac{\Gamma(\hat{P}(2),\hat{P}(3))-\Gamma(\hat{P}(1),\hat{P}(2))}{\Gamma(\hat{P}(2),\hat{P}(3))-\Gamma(\hat{P}(0),\hat{P}(1))}\right)\widehat{LATE}_{1}^{IV}\right],
\]
for $\xi\in(0,1)$. To maximize precision, one can set $\xi=[\hat{v}_{2}-\hat{v}_{12}]/[\hat{v}_{1}+\hat{v}_{2}-2\hat{v}_{12}]$
, where $\hat{v}_{1}$ and $\hat{v}_{2}$ are estimated variances
of $\widehat{LATE}_{2}^{IV}$ and the term in brackets, respectively,
and $\hat{v}_{12}$ is their covariance. By construction, $\widehat{LATE}{}_{2}^{\xi}$
provides an estimate of $LATE_{2}$ more precise than $\widehat{LATE}_{2}^{IV}$.
Though $\widehat{LATE}_{2}^{\xi}$ will tend to be less precise than
$\widehat{LATE}_{2}^{CF}$ when restriction (\ref{eq:multi_z-1})
is true, the probability limit of $\widehat{LATE}_{2}^{\xi}$ retains
an interpretation as a weighted average of causal effects for complier
subpopulations when (\ref{eq:multi_z-1}) is violated, a robustness
property emphasized elsewhere by \citet{angristpischke2009}.

\section{Equivalence failures}

Though Theorem \ref{thm:LATE_equivalence} establishes equivalence
between IV and a wide class of control function estimates of LATE,
other control function estimators fail to match IV even with a single
binary instrument. \citet{lalonde_1986} considered OLS estimation
of the following model:
\begin{equation}
Y_{i}=\alpha+\beta D_{i}+\gamma\left[D_{i}\times\left(-\tfrac{\phi(\Phi^{-1}(\hat{P}(Z_{i})))}{\hat{P}(Z_{i})}\right)+(1-D_{i})\times\left(\tfrac{\phi(\Phi^{-1}(\hat{P}(Z_{i})))}{1-\hat{P}(Z_{i})}\right)\right]+\epsilon_{i}.\label{eq:lalonde_cf}
\end{equation}
By imposing a common coefficient $\gamma$ on the Mills ratio terms
for the treatment and control groups, this specification allows for
selection on levels but rules out selection on treatment effects.

The term in brackets in equation (\ref{eq:lalonde_cf}) simplifies
to $(D_{i}-\hat{P}(Z_{i}))\times\{-\phi(\Phi^{-1}(\hat{P}(Z_{i})))/[\hat{P}(Z_{i})(1-\hat{P}(Z_{i}))]\}$.
When $\hat{P}(1)=1-\hat{P}(0)$ this term is proportional to the first
stage residual and least squares estimation of (\ref{eq:lalonde_cf})
yields an estimate of $\beta$ numerically identical to IV. This is
a finite sample analogue of Heckman and Vytlacil's \citeyearpar{heckman_vytlacil_nber2000}
observation (elaborated upon in \citealp{angrist_2004_tehet}) that
LATE equals ATE when both the first stage and the error distribution
are symmetric. When $\hat{P}(1)\neq1-\hat{P}(0)$, however, the control
function in equation (\ref{eq:lalonde_cf}) differs from the first
stage residual and the estimate of $\beta$ will not match IV. 
\begin{rem}
\noindent When $\hat{P}(1)=1-\hat{P}(0)$, the ATE estimate $\hat{\alpha}_{1}-\hat{\alpha}_{0}$
from a control function estimator of the form given in (\ref{eq:second_step})
coincides with IV whenever $J(U_{i})$ is presumed to follow a symmetric
distribution.
\end{rem}

\subsection*{Moments and monotonicity}

Theorem \ref{thm:LATE_equivalence} relied upon the fact that equation
(\ref{eq:linear}) includes enough free parameters to allow the control
function estimator to match the sample mean of $Y_{i}$ for every
combination of $D_{i}$ and $Z_{i}$. One might be tempted to conclude
that any structural estimator that fits these moments will produce
a corresponding LATE estimate equal to IV. We now show that this is
not the case. 

Suppose that treatment status is generated by a heterogeneous threshold
crossing model:
\begin{equation}
D_{i}=1\left\{ \kappa+\delta_{i}Z_{i}\geq U_{i}\right\} ,\label{eq:humpty}
\end{equation}

\noindent where $U_{i}$ is uniformly distributed and the random coefficient
$\delta_{i}$ is a mixture taking values in $\{-\eta,\eta\}$ for
some known positive constant $\eta$. Define $\upsilon\equiv Pr\left[\delta_{i}=\eta\right]$,
and suppose that $\delta_{i}$ is independent of $(Y_{i}(1),Y_{i}(0),U_{i},Z_{i})$.
Note that this model does not admit a representation of the form of
equation (\ref{eq:choice}) as it allows $D_{i}(1)<D_{i}(0)$.

Model (\ref{eq:humpty}) has two unknown parameters, $\kappa$ and
$\upsilon$, and can therefore rationalize the two observed choice
probabilities by choosing $\hat{\kappa}=\hat{P}(0)$ and $\hat{\upsilon}=(\eta+\hat{P}(1)-\hat{P}(0))/2\eta$.
Equations (\ref{eq:linear}) and (\ref{eq:humpty}) imply

\begin{center}
$E\left[Y_{i}|D_{i}=d,Z_{i}\right]=\alpha_{d}+\gamma_{d}\times\left[\upsilon\lambda_{d}\left(\kappa+\eta Z_{i}\right)+(1-\upsilon)\lambda_{d}(\kappa-\eta Z_{i})\right]$.
\par\end{center}

\noindent As before, we can use $\hat{\kappa}$ and $\hat{\upsilon}$
to construct control functions to include in a second-step regression,
producing estimates $\hat{\alpha}_{d}$ and $\hat{\gamma}_{d}$ that
exactly fit $\bar{Y}_{d}^{1}$ and $\bar{Y}_{d}^{0}$. 

Though this estimator matches all choice probabilities and conditional
mean outcomes, it produces an estimate of LATE different from IV.
The model's implied LATE is

\begin{center}
$E\left[Y_{i}(1)-Y_{i}(0)|D_{i}(1)>D_{i}(0)\right]=(\alpha_{1}-\alpha_{0})+(\gamma_{1}-\gamma_{0})\times E\left[J(U_{i})-\mu_{J}|\delta_{i}=\eta,\kappa<U_{i}\leq\kappa+\eta\right]$.
\par\end{center}

\noindent The corresponding control function estimator of this quantity
is
\begin{equation}
\widehat{LATE}^{*}=\left(\hat{\alpha}_{1}-\hat{\alpha}_{0}\right)+\left(\hat{\gamma}_{1}-\hat{\gamma}_{0}\right)\times\left(\tfrac{(\hat{\kappa}+\eta)\lambda_{1}(\hat{\kappa}+\eta)-\hat{\kappa}\lambda_{1}(\hat{\kappa})}{\eta}\right).\label{eq:fake_late}
\end{equation}

\noindent It is straightforward to verify that $\widehat{LATE}^{*}$
is not equal to $\widehat{LATE}^{IV}$. Equivalence fails here because
the selection model implies the presence of ``defiers'' with $D_{i}(1)<D_{i}(0)$.
IV does not identify LATE when there are defiers; hence, the model
suggests using a different function of the data to estimate the LATE.

\subsection*{Covariates}

\noindent It is common to condition on a vector of covariates $X_{i}$
either to account for possible violations of the exclusion restriction
or to increase precision. Theorem \ref{thm:LATE_equivalence} implies
that IV and control function estimates of LATE coincide if computed
separately for each value of the covariates, but this may be impractical
or impossible when $X_{i}$ can take on many values. 

A standard approach to introducing covariates is to enter them additively
into the potential outcomes model (see, e.g., \citealp{cornelissen_review,kline_walters_2016};
and \citealp{brinch_etal}). Suppose treatment choice is given by
$D_{i}=1\{P(X_{i},Z_{i})\geq U_{i}\}$ with $U_{i}$ independent of
$(X_{i},Z_{i})$, and assume
\begin{equation}
E\left[Y_{i}(d)|U_{i},X_{i}\right]=\alpha_{d}+\gamma_{d}\times(J(U_{i})-\mu_{J})+X_{i}^{\prime}\tau,\ d\in\{0,1\}.\label{eq:Xlinear}
\end{equation}
Letting $\hat{P}(X_{i},Z_{i})$ denote an estimate of $Pr[D_{i}=1|X_{i},Z_{i}]$,
the control function estimates for this model are
\begin{equation}
\left(\hat{\alpha}_{1},\hat{\gamma}_{1},\hat{\alpha}_{0},\hat{\gamma}_{0},\hat{\tau}\right)={\displaystyle {\displaystyle \arg\min_{\alpha_{1},\gamma_{1},\alpha_{0},\gamma_{0},\tau}}\sum_{i}\sum_{d\in\{0,1\}}1\left\{ D_{i}=d\right\} \left[Y_{i}-\alpha_{d}-\gamma_{d}\lambda_{d}(\hat{P}(X_{i},Z_{i}))-X_{i}'\tau\right]^{2}}.\label{eq:cf_covs}
\end{equation}

To ease exposition, we will study the special case of a single binary
covariate $X_{i}\in\{0,1\}$. Define $LATE(x)\equiv E[Y_{i}(1)-Y_{i}(0)|P(x,0)<U_{i}\leq P(x,1),X_{i}=x]$
as the average treatment effect for compliers with $X_{i}=x$, and
let $\hat{\alpha}_{d}(x)$ and $\hat{\gamma}_{d}(x)$ denote estimates
from unrestricted control function estimation among the observations
with $X_{i}=x$. The additive separability restriction in (\ref{eq:Xlinear})
suggests the following two estimators of $LATE(1)$:
\[
\widehat{LATE}_{x}^{CF}(1)=\left(\hat{\alpha}_{1}(x)-\hat{\alpha}_{0}(x)\right)+\left(\hat{\gamma}_{1}(x)-\hat{\gamma}_{0}(x)\right)\Gamma(\hat{P}(1,0),\hat{P}(1,1)),\ x\in\{0,1\}.
\]

\noindent By Theorem \ref{thm:LATE_equivalence} $\widehat{LATE}_{1}^{CF}(1)$
is a Wald estimate for the $X_{i}=1$ sample. $\widehat{LATE}_{0}^{CF}(1)$
gives an estimated effect for compliers with $X_{i}=1$ based upon
control function estimates for observations with $X_{i}=0$. The following
proposition describes the relationship between these two estimators
and the restricted estimator of $LATE\left(1\right)$ based upon (\ref{eq:cf_covs}).
\begin{prop}
Suppose Conditions \ref{assu:1} and \ref{assu:2} hold for each value
of $X_{i}\in\{0,1\}$ and let $\widehat{LATE}_{r}^{CF}(1)=(\hat{\alpha}_{1}-\hat{\alpha}_{0})+(\hat{\gamma}_{1}-\hat{\gamma}_{0})\Gamma(\hat{P}(1,0),\hat{P}(1,1))$
denote an estimate of $LATE(1)$ based on (\ref{eq:cf_covs}). Then\label{prop:cov_formula}
\end{prop}
\begin{center}
$\widehat{LATE}_{r}^{CF}(1)=w\widehat{LATE}_{1}^{CF}(1)+(1-w)\widehat{LATE}_{0}^{CF}(1)+b_{1}\left(\hat{\gamma}_{1}(1)-\hat{\gamma}_{1}(0)\right)+b_{0}\left(\hat{\gamma}_{0}(1)-\hat{\gamma}_{0}(0)\right)$. 
\par\end{center}

\noindent \emph{The coefficients $w$, $b_{1}$, and $b_{0}$ depend
only on the joint empirical distribution of $D_{i}$, $X_{i}$, and
$\hat{P}(X_{i},Z_{i})$.}

\noindent \textbf{Proof:} See the Appendix. $\blacksquare$

\begin{rem}
Proposition \ref{prop:cov_formula} demonstrates that control function
estimation under additive separability gives a linear combination
of covariate-specific estimates plus terms that equal zero when the
separability restrictions hold exactly in the sample. One can show
that the coefficient $w$ need not lie between 0 and 1. By contrast,
two-stage least squares estimation of a linear model with an additive
binary covariate using all interactions of $X_{i}$ and $Z_{i}$ as
instruments generates a weighted average of covariate-specific IV
estimates \citep{angristpischke2009}.
\end{rem}

\begin{rem}
Consider the following extension of equation (\ref{eq:Xlinear}):
\end{rem}
\begin{center}
$E\left[Y_{i}(d)|U_{i},X_{i}\right]=\alpha_{d}+\gamma_{d}\times(J(U_{i})-\mu_{J})+X_{i}^{\prime}\tau_{dc}+1\{U_{i}\leq P(X_{i},0)\}X_{i}^{\prime}\tau_{at}+1\{U_{i}>P(X_{i},1)\}X_{i}^{\prime}\tau_{nt},\ d\in\{0,1\}$.
\par\end{center}

\noindent This equation allows different coefficients on $X_{i}$
for always takers, never takers, and compliers by interacting $X_{i}$
with indicators for thresholds of $U_{i}$, and also allows the complier
coefficients to differ for treated and untreated outcomes. When $X_{i}$
includes a mutually exclusive and exhaustive set of indicator variables
and $\hat{P}(X_{i},Z_{i})$ equals the sample mean of $D_{i}$ for
each $(X_{i},Z_{i})$, control function estimation of this model produces
the same estimate of $E[Y_{i}|X_{i},D_{i},D_{i}(1)>D_{i}(0)]$ as
the semi-parametric procedure of \citet{abadie_2003}. Otherwise the
estimates may differ even asymptotically as the control function estimator
employs a different set of approximation weights when the model is
misspecified. 

\begin{rem}
A convenient means of adjusting for covariates that maintains the
numerical equivalence of IV and control function estimates is to weight
each observation by $\omega_{i}=Z_{i}/\hat{e}(X_{i})+(1-Z_{i})/(1-\hat{e}(X_{i}))$
where $\hat{e}(x)\in\left(0,1\right)$ is a first step estimate of
$Pr\left[Z_{i}=1|X_{i}=x\right]$. It is straightforward to show that
the $\omega_{i}-$weighted IV and control function estimates of the
unconditional LATE will be identical, regardless of the propensity
score estimator $\hat{e}(X_{i})$ employed. See \citet{hull_jmp}
for a recent application of this approach to covariate adjustment
of a selection model.
\end{rem}

\section{Maximum likelihood}

A fully parametric alternative to two-step control function estimation
is to specify a joint distribution for the model's unobservables and
estimate the parameters in one step via full information maximum likelihood
(FIML). Consider a model that combines (\ref{eq:choice}) and (\ref{eq:independence})
with the distributional assumption
\begin{equation}
Y_{i}(d)|U_{i}\sim F_{Y|U}\left(y|U_{i};\theta_{d}\right),\label{eq:parametric}
\end{equation}

\noindent where $F_{Y|U}(y|u;\theta)$ is a conditional CDF indexed
by a finite dimensional parameter vector $\theta$. For example, a
fully parametric version of the Heckit model is $Y_{i}(d)|U_{i}\sim N\left(\alpha_{d}+\gamma_{d}\Phi^{-1}(U_{i}),\sigma_{d}^{2}\right)$.
Since the marginal distribution of $U_{i}$ is also known, this model
provides a complete description of the joint distribution of $(Y_{i}(d),U_{i})$.
FIML exploits this distributional knowledge, estimating the model's
parameters as 
\begin{equation}
\begin{aligned}\left(\hat{P}(0)^{ML},\hat{P}(1)^{ML},\hat{\theta}_{0}^{ML},\hat{\theta}_{1}^{ML}\right)=\arg & {\displaystyle \max_{(P(0),P(1),\theta_{0},\theta_{1})}}{\displaystyle \sum_{i}}D_{i}\log\left(\int_{0}^{P(Z_{i})}f_{Y|U}\left(Y_{i}|u;\theta_{1}\right)du\right)\\
 & +{\displaystyle \sum_{i}}(1-D_{i})\log\left(\int_{P(Z_{i})}^{1}f_{Y|U}(Y_{i}|u;\theta_{0})du\right),
\end{aligned}
\label{eq:ml}
\end{equation}

\noindent where $f_{Y|U}(\cdot|u;\theta_{d})\equiv dF_{Y|U}\left(.|u;\theta_{d}\right)$
denotes the density (or probability mass function) of $Y_{i}(d)$
given $U_{i}=u$. The corresponding FIML estimates of treated and
untreated complier means are

\begin{center}
$\hat{\mu}_{dc}^{ML}=\dfrac{\int_{\hat{P}(0)^{ML}}^{\hat{P}(1)^{ML}}\int_{-\infty}^{\infty}yf_{Y|U}(y|u;\hat{\theta}_{d}^{ML})dydu}{\hat{P}(1)^{ML}-\hat{P}(0)^{ML}}$,
\par\end{center}

\noindent and the FIML estimate of LATE is $\widehat{LATE}^{ML}=\hat{\mu}_{1c}^{ML}-\hat{\mu}_{0c}^{ML}$. 

\subsection*{Binary outcomes}

We illustrate the relationship between FIML and IV estimates of LATE
with the special case of a binary $Y_{i}$. A parametric model for
this setting is given by
\begin{equation}
\begin{aligned}Y_{i}(d)= & 1\left\{ \alpha_{d}\geq\epsilon_{id}\right\} ,\\
\epsilon_{id}|U_{i} & \sim F_{\epsilon|U}\left(\epsilon|U_{i};\rho_{d}\right),
\end{aligned}
\label{eq:ml_binary}
\end{equation}

\noindent where $F_{\epsilon|U}(\epsilon|u;\rho)$ is a conditional
CDF characterized by the single parameter $\rho$. Equations (\ref{eq:choice})
and (\ref{eq:ml_binary}) include six parameters, which matches the
number of observed linearly independent probabilities (two values
of $Pr\left[D_{i}=1|Z_{i}\right]$, and four values of $Pr\left[Y_{i}=1|D_{i},Z_{i}\right]$).
The model is therefore ``saturated'' in the sense that a model with
more parameters would be under-identified.

The following result establishes the conditions under which maximum
likelihood estimates of complier means (and therefore LATE) coincide
with IV.
\begin{prop}
\noindent Consider the model defined by (\ref{eq:choice}), (\ref{eq:independence})
and (\ref{eq:ml_binary}). Suppose that Conditions \ref{assu:1} and
\ref{assu:2} hold, and that the maximum likelihood problem (\ref{eq:ml})
has a unique solution. Then $\hat{\mu}_{dc}^{ML}=\hat{\mu}_{dc}^{IV}$
for $d\in\{0,1\}$ if and only if $\hat{\mu}_{dc}^{IV}\in[0,1]$ for
$d\in\{0,1\}$.\label{prop:MLE_equivalence}
\end{prop}
\noindent \textbf{Proof:} See the Appendix. $\blacksquare$
\begin{rem}
The intuition for Proposition \ref{prop:MLE_equivalence} is that
the maximum likelihood estimation problem can be rewritten in terms
of the six identified parameters of the LATE model: $(\mu_{1at},\mu_{0nt},\mu_{1c},\mu_{0c},\pi_{at},\pi_{c})$,
where $\pi_{g}$ is the population share of group $g$. Unlike the
IV and control function estimators, the FIML estimator accounts for
the binary nature of $Y_{i}(d)$ by constraining all probabilities
to lie in the unit interval. When these constraints do not bind the
FIML estimates coincide with nonparametric IV estimates, but the estimates
differ when the nonparametric approach produces complier mean potential
outcomes outside the logically possible bounds. Logical violations
of this sort have been proposed elsewhere as a sign of failure of
instrument validity \citep{balke_pearle,imbens_rubin_97,huber_mellace,kitagawa_2015}.
\end{rem}

\begin{rem}
A simple ``limited information'' approach to maximum likelihood
estimation is to estimate $P\left(0\right)$ and $P\left(1\right)$
in a first step and then maximize the plug-in conditional log-likelihood
function
\end{rem}
\begin{center}
${\displaystyle \sum_{i}}D_{i}\log\left(\int_{0}^{\hat{P}(Z_{i})}f_{Y|U}\left(Y_{i}|u;\theta_{1}\right)du\right)+{\displaystyle \sum_{i}}(1-D_{i})\log\left(\int_{\hat{P}(Z_{i})}^{1}f_{Y|U}(Y_{i}|u;\theta_{0})du\right)$ 
\par\end{center}

\noindent with respect to $\left(\theta_{0},\theta_{1}\right)$ in
a second stage. One can show that applying this less efficient estimator
to a saturated model will produce an estimate of LATE equivalent to
IV under Conditions \ref{assu:1} and \ref{assu:2}. This broader
domain of equivalence results from some cross-equation parameter restrictions
being ignored by the two-step procedure. For example, the FIML estimator
may choose an estimate of $\pi_{c}$ other than $\hat{P}(1)-\hat{P}(0)$
in order to enforce the constraint that $(\mu_{1c},\mu_{0c})\in[0,1]^{2}$.

\subsection*{Overidentified models}

Equivalence of FIML and IV estimates at interior solutions in our
binary example follows from the fact that the model satisfies monotonicity
and includes enough parameters to match all observed choice probabilities.
Similar arguments apply to FIML estimators of sufficiently flexible
models for multi-valued outcomes. When the model includes fewer parameters
than observed choice probabilities, overidentification ensues. For
example, the standard bivariate probit model is a special case of
(\ref{eq:ml_binary}) that uses a normal distribution for $F_{\epsilon|U}(\cdot)$
and imposes $\epsilon_{i1}=\epsilon_{i0}$ and therefore $\rho_{1}=\rho_{0}$
(see \citealp{greene_text}). Hence, only five parameters are available
to rationalize six linearly independent probabilities. 

Maximum likelihood estimation of this more parsimonious model may
yield an estimate of LATE that differs from IV even at interior solutions.
This divergence stems from the model's overidentifying restrictions
which, if correct, may yield efficiency gains but if wrong can compromise
consistency. Though maximum likelihood estimation of misspecified
models yields a global best approximation to the choice probabilities
\citep{white_1982_mle}, there is no guarantee that it will deliver
a particularly good approximation to the LATE.

\section{Model evaluation}

In practice researchers often estimate selection models that impose
additive separability assumptions on exogenous covariates, combine
multiple instruments, and employ additional smoothness restrictions
that break the algebraic equivalence of structural LATE estimates
with IV. The equivalence results developed above provide a useful
conceptual benchmark for assessing the performance of structural models
in such applications. An estimator derived from a properly specified
model of treatment assignment and potential outcomes should come close
to matching a nonparametric IV estimate of the same parameter. Significant
divergence between these estimates would signal that the restrictions
imposed by the structural model are violated.

Figure 3 shows an example of this approach to model assessment from
Kline and Walters' \citeyearpar{kline_walters_2016} reanalysis of
the Head Start Impact Study (HSIS) \textendash{} a randomized experiment
with two-sided non-compliance \citep{puma_hsis}. On the vertical
axis are non-parametric IV estimates of the LATE associated with participating
in the Head Start program relative to a next best alternative for
various subgroups in the HSIS defined by experimental sites and baseline
child and parent characteristics. On the horizontal axis are two-step
control function estimates of the same parameters derived from a heavily
over-identified selection model involving multiple endogenous variables,
baseline covariates, and excluded instruments. Had this model been
saturated, all of the points would lie on the 45 degree line. In fact,
a Wald test indicates these deviations from the 45 degree line cannot
be distinguished from noise at conventional significance levels, suggesting
that the approximating model is not too far from the truth.
\begin{center}
\includegraphics[scale=0.77]{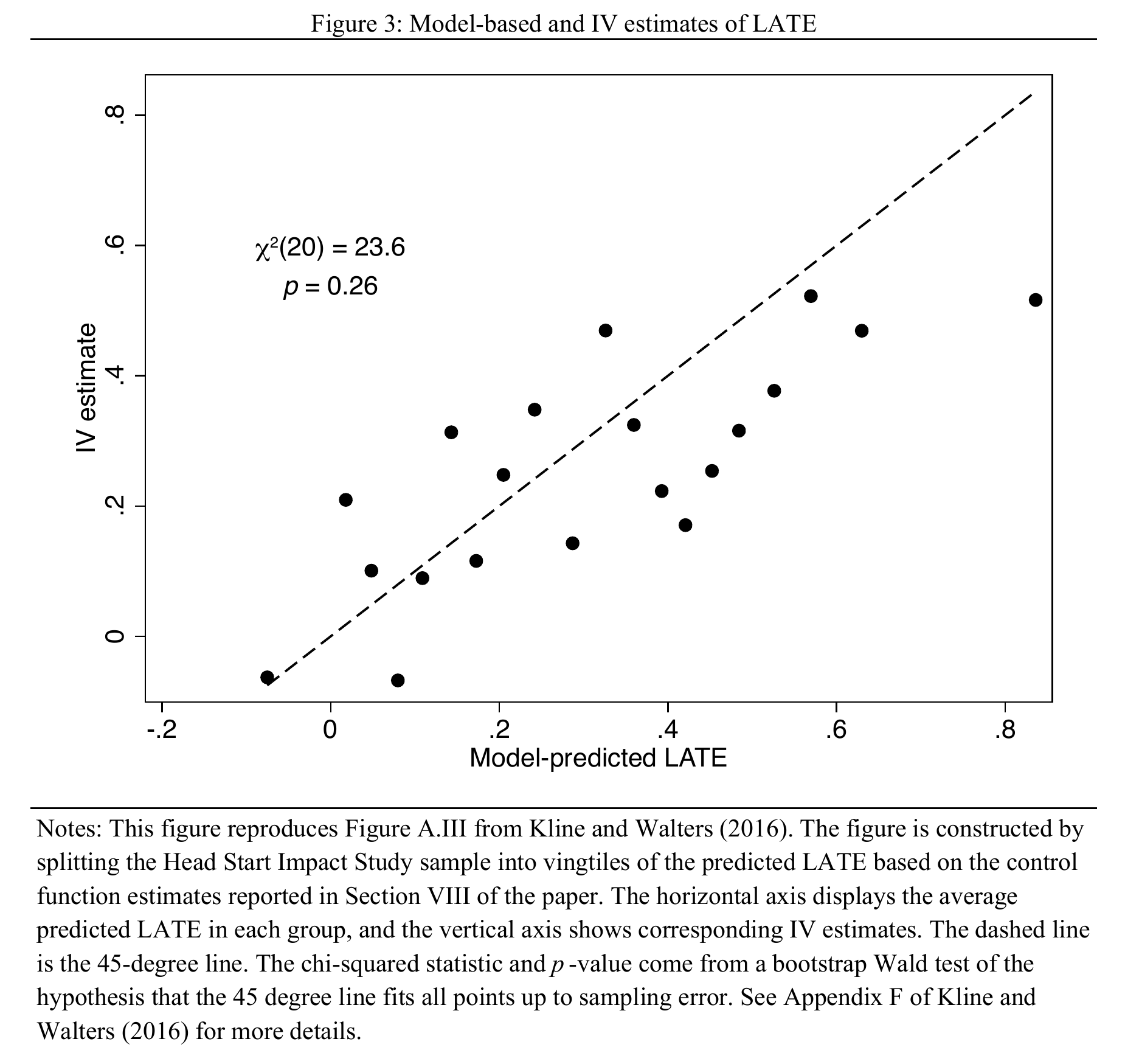}
\par\end{center}

Passing a specification test does not obviate the fundamental identification
issues inherent in interpolation and extrapolation exercises. As philosophers
of science have long argued, however, models that survive empirical
scrutiny deserve greater consideration then those that do not \citep{popper_1959,lakatos_1976}.
Demonstrating that a tightly restricted model yields a good fit to
IV estimates not only bolsters the credibility of the model's counterfactual
predictions, but serves to clarify what the estimated structural parameters
have to say about the effects of a research design as implemented.
Here the control function estimates reveal that Head Start had very
different effects on different sorts of complying households, a finding
rationalized by estimated heterogeneity in both patterns of selection
into treatment and potential outcome distributions.

\section{Conclusion}

This paper shows that two-step control function estimators of LATE
derived from a wide class of parametric selection models coincide
with the instrumental variables estimator. Control function and IV
estimates of mean potential outcomes for compliers, always takers,
and never takers are also equivalent. While many parametric estimators
produce the same estimate of LATE, different parameterizations can
produce dramatically different estimates of population average treatment
effects and other under-identified quantities. The sensitivity of
average treatment effect estimates to the choice of functional form
may be the source of the folk wisdom that structural estimators are
less robust than instrumental variables estimators. Our results show
that this view confuses robustness for a given target parameter with
the choice of target parameter. 

Structural estimators that impose overidentifying restrictions may
generate LATE estimates different from IV. Reporting the LATEs implied
by such estimators facilitates comparisons with unrestricted IV estimates
and is analogous to the standard practice of reporting average marginal
effects in binary choice models \citep{wooldridge_text}. Such comparisons
provide a convenient tool for assessing the behavioral restrictions
imposed by structural models. Model-based estimators that cannot rationalize
unrestricted IV estimates of LATE are unlikely to fare much better
at extrapolating to fundamentally under-identified quantities. On
the other hand, a tightly constrained structural estimator that fits
a collection of disparate IV estimates enjoys some degree of validation
that bolsters the credibility of its counterfactual predictions.

\bibliographystyle{ecta}
\bibliography{cw}
\pagebreak{}

\section*{Appendix }

\subsection*{Proof of Proposition \ref{prop:multi_z_equivalence}}

We begin by rewriting the IV and control function estimates of $LATE_{z}$
in matrix form. The IV estimator is given by

\begin{center}
$\widehat{LATE}_{z}^{IV}=\dfrac{\left[\hat{P}(z)\bar{Y}_{1}^{z}+(1-\hat{P}(z))\bar{Y}_{0}^{z}\right]-\left[\hat{P}(z-1)\bar{Y}_{1}^{z-1}+(1-\hat{P}(z-1))\bar{Y}_{0}^{z-1}\right]}{\hat{P}(z)-\hat{P}(z-1)}$
\par\end{center}

\begin{center}
$=\Psi_{1}^{z}(\hat{P})^{\prime}\bar{\mathbf{Y}}_{1}-\Psi_{0}^{z}(\hat{P})^{\prime}\bar{\mathbf{Y}}_{0},$
\par\end{center}

\noindent where $\bar{\mathbf{Y}}_{d}\equiv(\bar{Y}_{d}^{0},\bar{Y}_{d}^{1},....,\bar{Y}_{d}^{K})^{\prime}$
is the $(K+1)\times1$ vector of sample average outcomes for each
value of $z$ conditional on $D_{i}=d$ and $\hat{P}$ is the vector
of propensity score estimates. The $(K+1)\times1$ vector $\Psi_{1}^{z}(\hat{P})$
has $-\hat{P}(z-1)/[\hat{P}(z)-\hat{P}(z-1)]$ at entry $z-1$, $\hat{P}(z)/[\hat{P}(z)-\hat{P}(z-1)]$
at entry $z$, and zeros elsewhere, and the $(K+1)\times1$ vector
$\Psi_{0}^{z}(\hat{P})$ has $(1-\hat{P}(z-1))/[\hat{P}(z)-\hat{P}(z-1)]$
at entry $z-1$, $-(1-\hat{P}(z))/[\hat{P}(z)-\hat{P}(z-1)]$ at entry
$z$, and zeros elsewhere:

\begin{center}
$\Psi_{1}^{z}(\hat{P})=\left(0,..0,\tfrac{-\hat{P}(z-1)}{\hat{P}(z)-\hat{P}(z-1)},\tfrac{\hat{P}(z)}{\hat{P}(z)-\hat{P}(z-1)},0,...,0\right)^{\prime}$,
\par\end{center}

\begin{center}
$\Psi_{0}^{z}(\hat{P})=\left(0,..0,\tfrac{(1-\hat{P}(z-1))}{\hat{P}(z)-\hat{P}(z-1)},\tfrac{-(1-\hat{P}(z))}{\hat{P}(z)-\hat{P}(z-1)},0,...,0\right)^{\prime}$.
\par\end{center}

The second-step control function estimates with $L=K$ can be rewritten

\begin{center}
$(\hat{\alpha}_{d},\hat{\gamma}_{d1},...,\hat{\gamma}_{dK})={\displaystyle \arg\min_{\alpha_{d},\gamma_{d1},...,\gamma_{dK}}}\sum_{i}1\{D_{i}=d\}\left[Y_{i}-\alpha_{d}-\sum_{\ell=1}^{K}1\{Z_{i}=z\}\gamma_{d\ell}\lambda_{d\ell}(\hat{P}(z))\right]^{2}$.
\par\end{center}

\noindent This is a saturated OLS regression of $Y_{i}$ on $Z_{i}$
for each treatment category. The coefficient estimates satisfy

\begin{center}
$\hat{\alpha}_{d}+{\displaystyle \sum}_{\ell=1}^{K}\hat{\gamma}_{d\ell}\lambda_{d\ell}(\hat{P}(z))=\bar{Y}_{d}^{z},\ d\in\{0,1\},\ z\in\{0,1,...K\}$.
\par\end{center}

\noindent Letting $\hat{\Delta}_{d}=(\hat{\alpha}_{d},\hat{\gamma}_{d1},...,\hat{\gamma}_{dK})^{\prime}$
denote the control function estimates for treatment value $d$, we
can write this system in matrix form as

\begin{center}
$\Lambda_{d}(\hat{P})\hat{\Delta}_{d}=\bar{\mathbf{Y}}_{d}$,
\par\end{center}

\noindent where the matrix $\Lambda_{d}(\hat{P})$ has ones in its
first column and $\lambda_{dj-1}(\hat{P}(k))$ in row $k$ and column
$j>1$:

\begin{center}
$\Lambda_{d}(\hat{P})=\left[\begin{array}{cccc}
1 & \lambda_{d1}(\hat{P}(1)) & \cdots & \lambda_{dK}(\hat{P}(1))\\
1 & \lambda_{d1}(\hat{P}(2)) & \cdots & \lambda_{dK}(\hat{P}(2))\\
\vdots & \vdots & \ddots & \vdots\\
1 & \lambda_{d1}(\hat{P}(K)) & \cdots & \lambda_{dK}(\hat{P}(K))
\end{array}\right].$
\par\end{center}

\noindent The control function estimates are therefore given by

\begin{center}
$\hat{\Delta}_{d}=\Lambda_{d}(\hat{P})^{-1}\bar{\mathbf{Y}}_{d}$.
\par\end{center}

\noindent The values of $\lambda_{dk}(\hat{P}(z))$ are well-defined
for all $(k,z)$ whenever $0<\hat{P}(z)<1\ \forall z$, and $\Lambda_{d}(\hat{P})$
is full rank if $\hat{P}(z)\neq\hat{P}(z^{\prime})$ whenever $z\neq z^{\prime}$.
These requirements hold if Conditions \ref{assu:1} and \ref{assu:2}
are true for every pair of instrument values, so the matrix $\Lambda_{d}(\hat{P})$
is invertible under the conditions of Proposition \ref{prop:multi_z_equivalence}
and the control function estimate $\hat{\Delta}_{d}$ exists.

In matrix form, the control function estimate of $LATE_{z}$ is given
by

\begin{center}
$\widehat{LATE}_{z}^{CF}=\Upsilon^{z}(\hat{P})^{\prime}\left(\hat{\Delta}_{1}-\hat{\Delta}_{0}\right)$,
\par\end{center}

\noindent where the $(K+1)\times1$ vector $\Upsilon^{z}(\hat{P})$
has first entry equal to unity and $k$th entry $\Gamma_{k-1}(\hat{P}(z-1),\hat{P}(z))$
for $k>1$:

\begin{center}
$\Upsilon^{z}(\hat{P})=\left(1,\tfrac{\hat{P}(z)\lambda_{11}(\hat{P}(z))-\hat{P}(z-1)\lambda_{11}(\hat{P}(z-1))}{\hat{P}(z)-\hat{P}(z-1)},...,\tfrac{\hat{P}(z)\lambda_{1K}(\hat{P}(z))-\hat{P}(z-1)\lambda_{1K}(\hat{P}(z-1))}{\hat{P}(z)-\hat{P}(z-1)}\right)^{\prime}$.
\par\end{center}

\noindent Plugging in the formulas for $\hat{\Delta}_{1}$ and $\hat{\Delta}_{0}$
yields

\begin{center}
$\widehat{LATE}_{z}^{CF}=\Upsilon^{z}(\hat{P})^{\prime}\Lambda_{1}(\hat{P})^{-1}\bar{\mathbf{Y}}_{1}-\Upsilon^{z}(\hat{P})^{\prime}\Lambda_{0}(\hat{P})^{-1}\bar{\mathbf{Y}}_{0}$.
\par\end{center}

\noindent The IV and control functions are therefore identical if
$\Psi_{d}^{z}(\hat{P})^{\prime}=\Upsilon^{z}(\hat{P})^{\prime}\Lambda_{d}(\hat{P})^{-1}$
for $d\in\{0,1\}$, or equivalently, if $\Lambda_{d}(\hat{P})^{\prime}\Psi_{d}^{z}(\hat{P})=\Upsilon^{z}(\hat{P})$
for $d\in\{0,1\}$. 

For $d=1$, we have

\begin{center}
$\Lambda_{1}(\hat{P})^{\prime}\Psi_{1}^{z}(\hat{P})=\left(1,\tfrac{\hat{P}(z)\lambda_{11}(\hat{P}(z))-\hat{P}(z-1)\lambda_{11}(\hat{P}(z-1))}{\hat{P}(z)-\hat{P}(z-1)},...,\tfrac{\hat{P}(z)\lambda_{1K}(\hat{P}(z))-\hat{P}(z-1)\lambda_{1K}(\hat{P}(z-1))}{\hat{P}(z)-\hat{P}(z-1)}\right)^{\prime}$
\par\end{center}

\begin{center}
$=\Upsilon^{z}(\hat{P})$.
\par\end{center}

\noindent \begin{flushleft}
For $d=0$, we have
\par\end{flushleft}

\begin{center}
$\Lambda_{0}(\hat{P})^{\prime}\Psi_{0}^{z}(\hat{P})=\left(1,\tfrac{\lambda_{01}(\hat{P}(z-1))(1-\hat{P}(z-1))-\lambda_{01}(\hat{P}(z))(1-\hat{P}(z))}{\hat{P}(z)-\hat{P}(z-1)},...,\tfrac{\lambda_{01}(\hat{P}(z-1))(1-\hat{P}(z-1))-\lambda_{01}(\hat{P}(z))(1-\hat{P}(z))}{\hat{P}(z)-\hat{P}(z-1)}\right)^{\prime}$
\par\end{center}

\begin{center}
$=\left(1,\tfrac{\lambda_{11}(\hat{P}(z))\hat{P}(z)-\lambda_{11}(\hat{P}(z-1))\hat{P}(z-1)}{\hat{P}(z)-\hat{P}(z-1)},...,\tfrac{\lambda_{1K}(\hat{P}(z))\hat{P}(z)-\lambda_{1K}(\hat{P}(z-1))\hat{P}(z-1)}{\hat{P}(z)-\hat{P}(z-1)}\right)^{\prime}$
\par\end{center}

\begin{center}
$=\Upsilon^{z}(\hat{P})$,
\par\end{center}

\noindent where the second equality follows from the fact that $p^{\prime}\lambda_{1\ell}(p^{\prime})-p\lambda_{1\ell}(p)=(1-p)\lambda_{0\ell}(p)-(1-p^{\prime})\lambda_{0\ell}(p^{\prime})$
for any $p$, $p^{\prime}$ and $\ell$. This implies that $\widehat{LATE}_{z}^{IV}$
and $\widehat{LATE}_{z}^{CF}$ are equal to the same linear combination
of $\bar{\mathbf{Y}}_{1}$ and $\bar{\mathbf{Y}}_{0}$, so these estimates
are identical for any $z$. 

\subsection*{Proof of Proposition \ref{prop:cov_formula}}

The unrestricted control function estimates come from the regression

\begin{center}
$Y_{i}=\alpha_{0}(0)(1-D_{i})(1-X_{i})+\gamma_{0}(0)(1-D_{i})(1-X_{i})\lambda_{0}(\hat{P}(0,Z_{i}))$ 
\par\end{center}

\begin{center}
$+\alpha_{0}(1)(1-D_{i})X_{i}+\gamma_{0}(1)(1-D_{i})X_{i}\lambda_{0}(\hat{P}(1,Z_{i}))$ 
\par\end{center}

\begin{center}
$+\alpha_{1}(0)D_{i}(1-X_{i})+\gamma_{1}(0)D_{i}(1-X_{i})\lambda_{1}(\hat{P}(0,Z_{i}))$ 
\par\end{center}

\begin{center}
$+\alpha_{1}(1)D_{i}X_{i}+\gamma_{1}(1)D_{i}X_{i}\lambda_{1}(\hat{P}(1,Z_{i}))+\epsilon_{i}$.
\par\end{center}

\noindent We can write this equation in matrix form as

\begin{center}
$Y=W\Delta+\epsilon$,
\par\end{center}

\noindent where $W$ is the matrix of regressors and $\Delta=(\alpha_{0}(0),\gamma_{0}(0),\alpha_{0}(1),\gamma_{0}(1),\alpha_{1}(0),\gamma_{1}(0),\alpha_{1}(1),\gamma_{1}(1))^{\prime}$
collects the control function coefficients. Under the conditions of
Proposition \ref{prop:cov_formula}, $W^{\prime}W$ has full rank
and the unrestricted control function estimates are

\begin{center}
$\hat{\Delta}_{u}=(W^{\prime}W)^{-1}W^{\prime}Y$.
\par\end{center}

The estimator in equation (\ref{eq:cf_covs}) imposes three restrictions:
$\alpha_{1}(1)-\alpha_{1}(0)=\alpha_{0}(1)-\alpha_{0}(0)$, $\gamma_{1}(1)=\gamma_{1}(0)$,
and $\gamma_{0}(1)=\gamma_{0}(0)$. The resulting estimates can be
written

\begin{center}
$(\hat{\Delta}_{r},\hat{\varrho})={\displaystyle \arg\min_{\Delta,\varrho}}\ (Y-W\Delta)^{\prime}(Y-W\Delta)-\varrho C\Delta$,
\par\end{center}

\noindent where 

\begin{center}
$C=\left[\begin{array}{cccccccc}
-1 & 0 & 1 & 0 & 1 & 0 & -1 & 0\\
0 & 0 & 0 & 0 & 0 & -1 & 0 & 1\\
0 & -1 & 0 & 1 & 0 & 0 & 0 & 0
\end{array}\right]$
\par\end{center}

\noindent and $\varrho$ is a Lagrange multiplier. Then

\begin{center}
$\hat{\Delta}_{r}=\hat{\Delta}_{u}+(W^{\prime}W)^{-1}C^{\prime}\hat{\varrho}$,
\par\end{center}

\begin{center}
$\hat{\varrho}=-(C(W^{\prime}W)^{-1}C^{\prime})^{-1}C\hat{\Delta}_{u}$.
\par\end{center}

For any estimate $\hat{\Delta}$, the corresponding estimate of LATE
for compliers with $X_{i}=1$ is $\Upsilon(\hat{P})^{\prime}\hat{\Delta}$,
with

\begin{center}
$\Upsilon(\hat{P})=(0,0,-1,-\Gamma(\hat{P}(1,0),\hat{P}(1,1)),0,0,1,\Gamma(\hat{P}(1,0),\hat{P}(1,1)))^{\prime}$.
\par\end{center}

\noindent \begin{flushleft}
The restricted estimate of $LATE(1)$ is therefore
\par\end{flushleft}

\begin{center}
$\widehat{LATE}_{r}^{CF}(1)=\Upsilon(\hat{P})^{\prime}\left(\hat{\Delta}_{u}+(W^{\prime}W)^{-1}C^{\prime}\hat{\varrho}\right)$
\par\end{center}

\begin{center}
$=\widehat{LATE}_{1}^{CF}(1)+\Upsilon(\hat{P})^{\prime}(W^{\prime}W)^{-1}C^{\prime}\left(C(W^{\prime}W)^{-1}C^{\prime}\right)^{-1}\zeta$,
\par\end{center}

\noindent where $\zeta=-C\hat{\Delta}_{u}$ is the constraint vector
evaluated at the unrestricted estimates:

\begin{center}
$\zeta=\left([\hat{\alpha}_{0}(0)-\hat{\alpha}_{0}(1)]-[\hat{\alpha}_{1}(0)-\hat{\alpha}_{1}(1)],\hat{\gamma}_{1}(0)-\hat{\gamma}_{1}(1),\hat{\gamma}_{0}(0)-\hat{\gamma}_{0}(1)\right)^{\prime}$.
\par\end{center}

Write $\Omega=(W^{\prime}W)^{-1}C^{\prime}\left(C(W^{\prime}W)^{-1}C^{\prime}\right)^{-1}$,
and let $\nu_{k}$ denote the $3\times1$ vector equal to the transpose
of the $k$th row of $\Omega$. Using the fact that a scalar is equal
to its trace, we can then write the difference in restricted and unrestricted
LATE estimates as

\begin{center}
$\widehat{LATE}_{r}^{CF}(1)-\widehat{LATE}_{1}^{CF}(1)=tr\left(\Upsilon(\hat{P})^{\prime}\Omega\zeta\right)$
\par\end{center}

\begin{center}
$=tr\left(\Omega\zeta\Upsilon(\hat{P})^{\prime}\right)$
\par\end{center}

\begin{center}
$=\varphi^{\prime}\zeta$, 
\par\end{center}

\noindent where $\varphi=\nu_{7}-\nu_{3}+\Gamma(\hat{P}(1,0),\hat{P}(1,1))(\nu_{8}-\nu_{4})\equiv(\varphi_{1},\varphi_{2},\varphi_{3})^{\prime}$.
Then

\begin{center}
$\widehat{LATE}_{r}^{CF}(1)-\widehat{LATE}_{1}^{CF}(1)=\varphi_{1}\left([\hat{\alpha}_{0}(0)-\hat{\alpha}_{0}(1)]-[\hat{\alpha}_{1}(0)-\hat{\alpha}_{1}(1)]\right)+\varphi_{2}\left(\hat{\gamma}_{1}(0)-\hat{\gamma}_{1}(1)\right)+\varphi_{3}\left(\hat{\gamma}_{0}(0)-\hat{\gamma}_{0}(1)\right)$
\par\end{center}

\begin{center}
$=\varphi_{1}\left(\widehat{LATE}_{1}^{CF}(1)-\widehat{LATE}_{0}^{CF}(1)\right)+\left(\hat{\gamma}_{1}(0)-\hat{\gamma}_{1}(1)\right)\left(\varphi_{2}+\varphi_{1}\Gamma(\hat{P}(1,0),\hat{P}(1,1))\right)$
\par\end{center}

\begin{center}
$+(\hat{\gamma}_{0}(0)-\hat{\gamma}_{0}(1))\left(\varphi_{3}-\varphi_{1}\Gamma(\hat{P}(1,0),\hat{P}(1,1))\right).$
\par\end{center}

\noindent This implies

\begin{center}
$\widehat{LATE}_{r}^{CF}(1)=w\widehat{LATE}_{1}^{CF}(1)+(1-w)\widehat{LATE}_{0}^{CF}(1)+b_{1}\left(\hat{\gamma}_{1}(1)-\hat{\gamma}_{1}(0)\right)+b_{0}\left(\hat{\gamma}_{0}(1)-\hat{\gamma}_{0}(0)\right)$,
\par\end{center}

\noindent where $w=1+\varphi_{1}$, $b_{1}=-(\varphi_{2}+\varphi_{1}\Gamma(\hat{P}(1,0),\hat{P}(1,1)))$,
and $b_{0}=\varphi_{1}\Gamma(\hat{P}(1,0),\hat{P}(1,1))-\varphi_{3}$.
Furthermore, note that the elements of $\varphi$ only depend on sample
moments of $D_{i}$, $X_{i}$, and $\hat{P}(X_{i},Z_{i})$, so the
proposition follows.

\subsection*{Proof of Proposition \ref{prop:MLE_equivalence}}

The log likelihood function for model (\ref{eq:ml_binary}) is

\begin{center}
${\displaystyle \log\mathcal{L}\left(P(0),P(1),\alpha_{0},\alpha_{1},\rho_{0},\rho_{1}\right)}={\displaystyle \sum_{i}D_{i}\log\left(\int_{0}^{P(Z_{i})}\left[Y_{i}F_{\epsilon|U}(\alpha_{1}|u;\rho_{1})+(1-Y_{i})(1-F_{\epsilon|U}(\alpha_{1}|u;\rho_{1}))\right]du\right)}$
\par\end{center}

\begin{center}
$+{\displaystyle \sum_{i}(1-D_{i})\log\left(\int_{P(Z_{i})}^{1}\left[Y_{i}F_{\epsilon|U}(\alpha_{0}|u;\rho_{0})+(1-Y_{i})(1-F_{\epsilon|U}(\alpha_{0}|u;\rho_{0}))\right]du\right)}$.
\par\end{center}

\noindent We first rewrite this likelihood in terms of the six identified
parameters of the LATE model, which are given by

\begin{center}
$\pi_{at}=P(0)$,
\par\end{center}

\begin{center}
$\pi_{c}=P(1)-P(0)$, 
\par\end{center}

\begin{center}
$\mu_{1at}=\dfrac{\int_{0}^{P(0)}F_{\epsilon|U}(\alpha_{1}|u;\rho_{1})du}{P(0)}$,
\par\end{center}

\begin{center}
$\mu_{0nt}=\dfrac{\int_{P(1)}^{1}F_{\epsilon|U}\left(\alpha_{0}|u;\rho_{0}\right)du}{1-P(1)},$
\par\end{center}

\begin{center}
$\mu_{dc}=\dfrac{\int_{P(0)}^{P(1)}F_{\epsilon|U}(\alpha_{d}|u;\rho_{d})du}{P(1)-P(0)},\ d\in\{0,1\}$. 
\par\end{center}

\noindent Note that since $F_{\epsilon|U}(\cdot|u;\rho)$ is a CDF,
we have $\mu_{dg}\in[0,1]\ \forall(d,g)$. Substituting these parameters
into the likelihood function yields

\begin{center}
${\displaystyle \log\mathcal{L}(\pi_{at},\pi_{c},\mu_{1at},\mu_{0nt},\mu_{1c},\mu_{0c})=}{\displaystyle \sum_{i}D_{i}Z_{i}\log\left(\pi_{at}\left[Y_{i}\mu_{1at}+(1-Y_{i})(1-\mu_{1at})\right]+\pi_{c}\left[Y_{i}\mu_{1c}+(1-Y_{i})(1-\mu_{1c})\right]\right)}$
\par\end{center}

\begin{center}
$+{\displaystyle \sum_{i}D_{i}(1-Z_{i})\log\left(\pi_{at}\left[Y_{i}\mu_{1at}+(1-Y_{i})(1-\mu_{1at})\right]\right)}$
\par\end{center}

\begin{center}
$+{\displaystyle \sum_{i}(1-D_{i})Z_{i}\log\left((1-\pi_{at}-\pi_{c})\left[Y_{i}\mu_{0nt}+(1-Y_{i})(1-\mu_{0nt})\right]\right)}$
\par\end{center}

\begin{center}
$+{\displaystyle \sum_{i}(1-D_{i})(1-Z_{i})\log\left((1-\pi_{at}-\pi_{c})\left[Y_{i}\mu_{0nt}+(1-Y_{i})(1-\mu_{0nt})\right]+\pi_{c}\left[Y_{i}\mu_{0c}+(1-Y_{i})(1-\mu_{0c})\right]\right)}$.
\par\end{center}

We first consider interior solutions. The first-order conditions are

\begin{center}
$[\mu_{1at}]:\ {\displaystyle \sum_{i}}\left(\tfrac{D_{i}(2Y_{i}-1)Z_{i}\pi_{at}}{\pi_{at}\left[Y_{i}\mu_{1at}+(1-Y_{i})(1-\mu_{1at})\right]+\pi_{c}\left[Y_{i}\mu_{1c}+(1-Y_{i})(1-\mu_{1c})\right]}+{\displaystyle \tfrac{D_{i}(2Y_{i}-1)(1-Z_{i})\pi_{at}}{\pi_{at}\left[Y_{i}\mu_{1at}+(1-Y_{i})(1-\mu_{1at})\right]}}\right)=0$,
\par\end{center}

\begin{center}
$[\mu_{0nt}]:\ {\displaystyle \sum_{i}}\left(\tfrac{(1-D_{i})(2Y_{i}-1)Z_{i}(1-\pi_{at}-\pi_{c})}{(1-\pi_{at}-\pi_{c})\left[Y_{i}\mu_{0nt}+(1-Y_{i})(1-\mu_{0nt})\right]}+\tfrac{(1-D_{i})(2Y_{i}-1)(1-Z_{i})(1-\pi_{at}-\pi_{c})}{(1-\pi_{at}-\pi_{c})\left[Y_{i}\mu_{0nt}+(1-Y_{i})(1-\mu_{0nt})\right]+\pi_{c}\left[Y_{i}\mu_{0c}+(1-Y_{i})(1-\mu_{0c})\right]}\right)=0$,
\par\end{center}

\begin{center}
$[\mu_{1c}]:\ {\displaystyle \sum_{i}}\tfrac{D_{i}Z_{i}(2Y_{i}-1)\pi_{c}}{\pi_{at}\left[Y_{i}\mu_{1at}+(1-Y_{i})(1-\mu_{1at})\right]+\pi_{c}\left[Y_{i}\mu_{1c}+(1-Y_{i})(1-\mu_{1c})\right]}=0$,
\par\end{center}

\begin{center}
$[\mu_{0c}]:\ {\displaystyle \sum_{i}\tfrac{(1-D_{i})(1-Z_{i})(2Y_{i}-1)\pi_{c}}{(1-\pi_{at}-\pi_{c})\left[Y_{i}\mu_{0nt}+(1-Y_{i})(1-\mu_{0nt})\right]+\pi_{c}\left[Y_{i}\mu_{0c}+(1-Y_{i})(1-\mu_{0c})\right]}}=0$,
\par\end{center}

\begin{center}
$[\pi_{at}]:\ {\displaystyle \sum_{i}}\tfrac{D_{i}Z_{i}\left[Y_{i}\mu_{1at}+(1-Y_{i})(1-\mu_{1at})\right]}{\pi_{at}\left[Y_{i}\mu_{1at}+(1-Y_{i})(1-\mu_{1at})\right]+\pi_{c}\left[Y_{i}\mu_{1c}+(1-Y_{i})(1-\mu_{1c})\right]}+{\displaystyle \sum_{i}\tfrac{D_{i}(1-Z_{i})}{\pi_{at}}}$
\par\end{center}

\begin{center}
$-{\displaystyle \sum_{i}\tfrac{(1-D_{i})Z_{i}\left[Y_{i}\mu_{0nt}+(1-Y_{i})(1-\mu_{0nt})\right]}{(1-\pi_{at}-\pi_{c})\left[Y_{i}\mu_{0nt}+(1-Y_{i})(1-\mu_{0nt})\right]}}-{\displaystyle \sum_{i}\tfrac{(1-D_{i})(1-Z_{i})\left[Y_{i}\mu_{0nt}+(1-Y_{i})(1-\mu_{0nt})\right]}{(1-\pi_{at}-\pi_{c})\left[Y_{i}\mu_{0nt}+(1-Y_{i})(1-\mu_{0nt})\right]+\pi_{c}\left[Y_{i}\mu_{0c}+(1-Y_{i})(1-\mu_{0c})\right]}}=0$,
\par\end{center}

\begin{center}
$[\pi_{c}]:\ {\displaystyle \sum_{i}\tfrac{D_{i}Z_{i}\left[Y_{i}\mu_{1c}+(1-Y_{i})(1-\mu_{1c})\right]}{\left[\pi_{at}\left[Y_{i}\mu_{1at}+(1-Y_{i})(1-\mu_{1at})\right]+\pi_{c}\left[Y_{i}\mu_{1c}+(1-Y_{i})(1-\mu_{1c})\right]\right]}}-{\displaystyle \sum_{i}\tfrac{(1-D_{i})Z_{i}\left[Y_{i}\mu_{0nt}+(1-Y_{i})(1-\mu_{0nt})\right]}{(1-\pi_{at}-\pi_{c})\left[Y_{i}\mu_{0nt}+(1-Y_{i})(1-\mu_{0nt})\right]}}$
\par\end{center}

\begin{center}
$-{\displaystyle \sum_{i}\tfrac{(1-D_{i})(1-Z_{i})(2Y_{i}-1)(\mu_{0nt}-\mu_{0c})}{(1-\pi_{at}-\pi_{c})\left[Y_{i}\mu_{0nt}+(1-Y_{i})(1-\mu_{0nt})\right]+\pi_{c}\left[Y_{i}\mu_{0c}+(1-Y_{i})(1-\mu_{0c})\right]}}=0$. 
\par\end{center}

\noindent \begin{flushleft}
Under Conditions \ref{assu:1} and \ref{assu:2} we can compute $\hat{\mu}_{1at}^{IV}$,
$\hat{\mu}_{0nt}^{IV}$, $\hat{\mu}_{1c}^{IV}$, and $\hat{\mu}_{0c}^{IV}$
. Setting $\hat{\pi}_{c}^{IV}=\hat{P}(1)-\hat{P}(0)$ and $\hat{\pi}_{at}^{IV}=\hat{P}(0)$
and plugging the IV parameter estimates into the FIML first order
conditions, we see that these conditions are satisfied. Thus at interior
solutions maximum likelihood and IV estimators of all parameters are
equal, and it follows that $\hat{\mu}_{dc}^{ML}=\hat{\mu}_{dc}^{IV}$
for $d\in\{0,1\}$. 
\par\end{flushleft}

Next, we consider corner solutions, which occur when at least one
parameter lies outside {[}0,1{]} at the unconstrained solution to
the first order conditions. Note that $\widehat{\mu}_{1at}^{IV}$,
$\widehat{\mu}_{0nt}^{IV}$, and $\hat{\pi}_{at}^{IV}$ are sample
means of binary variables, so these estimates are always in the unit
interval. $\hat{\pi}_{c}^{IV}$ is the difference in empirical treatment
rates between the two values of $Z_{i}$; without loss of generality
we assume that $Z_{i}=1$ refers to the group with the higher treatment
rate, so $\hat{\pi}_{c}^{IV}\in(0,1)$. Thus a constraint binds if
and only if $\hat{\mu}_{dc}^{IV}$ is outside {[}0,1{]} for $d=0$,
$d=1$, or both. In these cases at least one of the maximum likelihood
complier means fails to match the corresponding IV estimate because
the IV estimate is outside the FIML parameter space. This establishes
that the FIML and IV estimates match if and only if both $\hat{\mu}_{1c}^{IV}$
and $\hat{\mu}_{0c}^{IV}$ are in {[}0,1{]}, which completes the proof.
\end{document}